\documentclass[11pt,a4paper]{article}
\pdfoutput=1
\usepackage{myjheppub}
\usepackage{latexsym,amsfonts,amsmath,amssymb}
\usepackage{amsthm}
\usepackage{mathtools}
\usepackage{bbm}
\usepackage{graphicx}
\usepackage{color}
\usepackage{caption}
\usepackage{subcaption}
\usepackage[normalem]{ulem}
\usepackage{comment}
\usepackage{url}
\usepackage{slashed}
\usepackage{bm}
\usepackage{hhline}
\usepackage{physics}

\usepackage{tabu}

\usepackage{footmisc}

\usepackage{tikz}
\usetikzlibrary{decorations.pathreplacing, decorations.markings,calc,shapes.misc,decorations.pathmorphing,patterns.meta, math}
\usetikzlibrary{arrows}

\newcommand{\CG}{\mathcal{G}}
\newcommand{\CK}{\mathcal{K}}

\newcommand{\CN}{\mathcal{N}}

\newcommand{\CY}{\mathcal{Y}}

\newcommand{\CD}{\mathcal{D}}

\newcommand\mx{{\mathbf{x}}}

\renewcommand{\Im}{{\rm Im}}
\renewcommand{\Re}{{\rm Re}}

\newcommand{\IR}{\mathbb{R}}

\newcommand{\SBHext}{S_{BH}^\text{extremal}}

\newcommand{\rme}{e}

\newcommand\be{\begin{equation}}
\newcommand\ee{\end{equation}}
\newcommand\bea{\begin{eqnarray}}
\newcommand\eea{\end{eqnarray}}

\renewcommand{\dd}{\mathrm{d}}

\renewcommand{\=}{\;= \;}

\renewcommand{\d}{\delta}
\renewcommand{\a}{\alpha}

\renewcommand{\t}{\tau}

\newcommand{\wt}{\widetilde}

\renewcommand{\Re}{\text{Re}}
\renewcommand{\Im}{\text{Im}}

\newcommand{\ve}{\varepsilon}

\newcommand{\p}{\partial}

\renewcommand{\i}{{\rm i}}

\newcommand{\half}{\frac12}

\newcommand{\x}{{\rm x}}

\newcommand{\SBH}{S_\text{BH}^\text{extremal}}

\newcommand{\bx}{\pmb \x}

\newcommand{\Hv}{H}
\newcommand{\tPhi}{\tilde{\Phi}}
\newcommand{\tphi}{\tilde{\phi}}
\newcommand{\tH}{\tilde{H}}

\newcommand{\nv}{n_\text{v}}

\renewcommand{\Im}{\mbox{Im}}
\renewcommand{\Re}{\mbox{Re}}

\newcommand{\hor}{\text{hor}}
\newcommand{\xvec}{\textbf{x}}
\newcommand{\omE}{\omega_E}
\newcommand{\omEphi}{\omega_{E \, \phi}}

\newcommand{\intprod}[2]{\langle \, #1 \,,\, #2\, \rangle}

\newcommand{\bast}{\boldsymbol{\ast}}

\title{New forms of attraction:\\ Attractor saddles for the black hole index 
}

\author{Jan Boruch$^1$, Luca Iliesiu${}^{2,3}$, Sameer Murthy$^{4,5}$, Gustavo J. Turiaci$^{6}$}

\affiliation{{${}^1$  Department of Physics, University of Warsaw, ul. Pasteura 5, 02-093 Warsaw, Poland}}
\affiliation{
{${}^2$ Stanford Institute for Theoretical Physics, Stanford University, Stanford, CA 94305, USA}}
\affiliation{{${}^3$ Department of Physics, University of California,
Berkeley, CA 94720, USA}}

\affiliation{${}^4$ Department of Mathematics, King’s College London, The Strand, London WC2R 2LS, UK}
   
\affiliation{${}^5$  School of Natural Sciences, Institute for Advanced Study, Princeton, NJ, USA}

\affiliation{${}^6$ Physics Department, University of Washington, Seattle, WA, USA}

\abstract{
The count of microstates for supersymmetric black holes is typically obtained from a supersymmetric index in weakly-coupled string theory. We find the saddles in the gravitational path integral corresponding to this index
in a general theory of $\CN=2$ supergravity in asymptotically flat space. This saddle exhibits a new attractor mechanism which explains the agreement between the string theory index and the macroscopic entropy. 
These saddles are smooth, complex Euclidean spinning black holes that are supersymmetric but not extremal, i.e.~they are formally finite-temperature solutions. With this new mechanism, the scalars and the electromagnetic fields get attracted to temperature- and moduli-independent values at the north and south poles of the rotating black hole, although they vary along the Euclidean horizon in a non-universal way.
Further, although the area and the spin of the black hole 
depend non-trivially on the temperature and on the moduli, the free energy 
is essentially a function only of the black hole charges (apart from a 
trivial dependence on the temperature and the moduli through the BPS mass), and agrees with the string theory index.
}

\begin{document}

\maketitle 

\section{Introduction}

The explanation of the thermodynamic entropy of black holes as the statistical entropy of an underlying ensemble of microstates 
is one of the big successes of string theory.
Supersymmetric black holes have played an important role in elucidating this idea~\cite{Sen:1995in,Strominger:1996sh}. 
For a large class of such black holes in string theory one can independently calculate the statistical and thermodynamic entropies 
and verify that they agree, see~\cite{Sen:2007qy} for a review. 
In some particularly symmetric situations, this match can be developed in great detail~\cite{Dabholkar:2011ec,Dabholkar:2014ema,Iliesiu:2022kny}.

\smallskip

In order to calculate the statistical entropy in string theory, one considers a supersymmetric compactification such as Type II string theory on a Calabi-Yau 3-fold. 
At weak coupling, one can identify the microstates as bound states of fluctuations of strings and branes, and enumerate them 
for a given set of charges. 
More precisely, one calculates a \emph{supersymmetric index} in the weakly-coupled string theory. 
Recall that the basic Witten index~\cite{Witten:1982df} is defined as a trace over the Hilbert space of the theory, 
\be \label{WitInd}
Z \= \text{Tr}_{\mathcal{H}} \, (-1)^{\sf F} \, e^{-\beta H} \= d^0_B-d^0_F  
\,, \qquad  H \= \{ Q, \overline{Q}\}\,.
\ee
Here~$(-1)^{\sf F}$ is the fermion number,  and the theory has a pair of complex supercharges 
that anticommute to the Hamiltonian~$H$ as shown. 
The trace is well-defined for any finite inverse temperature~$\beta>0$ because of the damping of the high-energy states. On the other hand, as is well-known, the supersymmetry algebra implies a pairing of states of non-zero energy, so 
that that index is independent of~$\beta$ and reduces to the difference~$d^0_B-d^0_F$ in the number of bosonic and fermionic states at zero energy. 
The index used in the enumeration of black hole microstates is typically some refinement of the Witten index, which produces an integer as in~\eqref{WitInd} that depends only the charges of the black hole. 

\smallskip

When the effective string coupling constant is strong, the theory is  described by supergravity coupled to  matter fields, 
and it admits a supersymmetric black hole solution carrying the same set of charges as the microstates.\footnote{The semiclassical gravitational solution as a saddle-point of the gravitational path integral is valid when the charges are large. In this paper we stay within this approximation.} 
As one approaches the horizon of the supersymmetric black hole, 
the geometry takes the form of 
a semi-infinite tube extending to the interior of spacetime. 
In the near-horizon region, the  geometry is fixed to be a product of AdS$_2$ and a compact manifold, 
and the electromagnetic field strengths and the scalar fields are all constant. 
Further, the values of all these fields are completely fixed in terms of the charges of the solution and, in particular, they are independent of the asymptotic values of the scalar fields (the \emph{moduli}).  
This phenomenon is referred to as the black hole \emph{attractor mechanism}~\cite{Ferrara:1995ih,Ferrara:1996dd}, and it explains, in particular, how the black hole entropy is a function only of the charges and independent of the moduli as for the statistical entropy.

\medskip

However, there is a basic tension between the attractor mechanism and the supersymmetric index. 
The main point is that the regular Lorentzian metric of supersymmetric black holes is necessarily extremal, 
as reflected in the emergence of the AdS$_2$ region.\footnote{Indeed, the near-horizon attractor configuration was 
shown to arise purely as a consequence of extremality (see e.g.~\cite{Sen:2005wa}).}  
The condition of extremality implies that~$\beta \to \infty$, whereas the index~\eqref{WitInd} is defined at finite~$\beta$. 
Moreover, there is no reason to expect a priori that the thermodynamic area of the extremal black hole, given by 
a quarter of its horizon area, includes the~$(-1)^{\sf F}$ insertion in the trace~\eqref{WitInd}.\footnote{We should emphasize that the agreement of the index and the entropy of supersymmetric black holes has been 
explained using a heuristic mixture of Lorentzian Hilbert space and Euclidean functional integral methods in the past~\cite{Sen:2009vz,Dabholkar:2010rm}. Further, 
these ideas have led to precise predictions which have been checked in detail~\cite{Manschot:2010qz,Manschot:2011xc, Sen:2011ktd, Bringmann:2012zr,Chattopadhyaya:2018xvg}. 
Here our aim is to explain the agreement of the index and entropy in one unified formalism.} 
One could imagine a possible resolution in the Euclidean signature where one has at least a semi-classical description of the trace as a functional integral in gravity. 
In the Euclidean solution, the size of the time circle in the asymptotic region becomes infinite. If one forces the asymptotic circle in this geometry to have a finite value of~$\beta$,  it creates a cusp in the interior \cite{Hawking:1994ii}. This geometry is problematic. For one, ad-hoc boundary conditions have to be imposed at the location of the cusp. 
Moreover, if one views this solution as a saddle-point contribution to the gravitational path integral its on-shell action predicts a zero entropy, in disagreement with both the string theory index and with the area of the extremal solution. 

The aim of this paper is to resolve this disagreement by finding the true gravitational saddle-point contributions to the index of attractor black holes.

\bigskip

We work in the context of the original attractor mechanism, namely four-dimensional $\CN=2$ 
supergravity coupled to a number of vector and hyper multiplets. 
This theory admits black hole solutions with magnetic and electric charges~$\Gamma=(P,Q)$ which are annihilated by four of the eight 
supercharges of the theory. The algebra of these supercharges in the~$\CN=2$ theory can be taken to be~$\{Q, \overline{Q} \} = E - |Z|$,
where~$E$ and~$Z$ are the energy and the central charge. 
The energy~$E$ of $\frac12$-BPS states is therefore determined in terms of the central charge to be~$|Z| \equiv M_\text{BPS}$. 
In this setting the relevant supersymmetric index to count the dual black hole microstates is the so-called helicity supertrace. 
After absorbing a certain number of fermion zero-modes, this reduces to the index, similar to \eqref{WitInd}\footnote{A decomposition of the total Hilbert space as $\mathcal{H} = \mathcal{H}_{\rm int} \otimes \mathcal{H}_{\rm CM}$ is usually assumed. $\mathcal{H}_{\rm int}$ is associated to the internal black hole microstates, while $\mathcal{H}_{\rm CM}$ to the center of mass degrees of freedom. Only in the semiclassical limit of the gravitational path integral, there is a clear distinction between the modes that correspond to each factor. If this decomposition were exact, the helicity supertrace on $\mathcal{H}$ reduces to the Witten index on $\mathcal{H}_{\rm int}$, namely $-2 {\rm Tr}_{\mathcal{H}} \, e^{2\pi \i J'} J'{}^2 = {\rm Tr}_{\mathcal{H}_{\rm int}} \, e^{2\pi \i J}$, where $J'$ is the total angular momentum in an arbitrary direction while $J$ is the black hole internal spin \cite{Dabholkar:2005by, Denef:2007vg}. The results here are a first step towards developing a formalism able to drop such assumptions.}, 
\be \label{eq:index}
Z_\Gamma (\beta) \= \text{Tr}_{\Gamma}  \, (-1)^{\sf F} \,  e^{-\beta E} \= 
\text{Tr}_{\Gamma}  \, e^{2 \pi \i J} \,  e^{-\beta \{ Q, \overline{Q}\} } \, e^{-\beta M_\text{BPS}}  
\= (d_B(\Gamma)- d_F(\Gamma)) \, e^{-\beta M_\text{BPS}} 
\,,
\ee
defined as a trace over the Hilbert subspace of fixed charge~$\Gamma$ in the microscopic string theory~\cite{Sen:2009vz,Dabholkar:2010rm}. To obtain the third equality we have used the anticommutator of the supercharges in the~$\CN=2$ theory and 
the spin-statistics theorem~$(-1)^{\sf F} = e^{2\pi \i J}$. As in~\eqref{WitInd}, the only surviving contributions to the trace come from BPS states and yields~$d_B(\Gamma)- d_F(\Gamma)$ which is the integer-valued index of BPS states. 

\smallskip

The main question we would like to answer is: what is the gravitational definition of the index~\eqref{eq:index}?
Suppose we consider the formal Euclidean path integral for quantum gravity, ignoring questions about the high-energy behavior, 
as in the Gibbons-Hawking approach. What are the saddles that contribute to this index? Is there a saddle that is related to the 
supersymmetric black hole, as suggested by the agreement of its entropy and the index? 
In this paper we find and elucidate the nature of such a  saddle point for the gravitational index in theories of four-dimensional~$\CN=2$ supergravity of the type that arise in Calabi-Yau compactifications of string theory.

To set up the gravitational path integral that computes the index, \eqref{eq:index} instructs us to not only make the time circle periodic but also set the angular velocity to be~$2\pi \i/\beta$. Saddle points with such boundary conditions were found in asymptotically AdS space~\cite{Cabo-Bizet:2018ehj}, and in asymptotically 
flatspace~\cite{Iliesiu:2021are}, 
thus making a direct link between the index and the gravitational path integral.\footnote{This was generalized to other asymptotic AdS spaces  in~\cite{Cassani:2019mms,Bobev:2019zmz,Larsen:2021wnu}. Such configurations were used in \cite{Heydeman:2020hhw,Boruch:2022tno} to calculate the detailed properties of the gravitational index, including the non-trivial fluctuations of the Schwarzian mode. 
The same procedure is even applicable without supersymmetry \cite{Chen:2023mbc}.} These saddles are smooth, supersymmetric, yet non-extremal, complex\footnote{Similar configurations were called ``quasi-Euclidean" in the Gibbons-Hawking formalism~\cite{Gibbons:1976ue}.} solutions of supergravity that exist solely because of their non-zero angular momentum. The regularized on-shell action of these cigar-like saddles is 
the free energy of the supersymmetric black hole and agrees with the index of the dual quantum theory.

\smallskip

In this paper, we generalize these ideas for the supersymmetric index in a theory of $\CN=2$ supergravity coupled to an arbitrary number of vector multiplets in four-dimensional flat space.
This theory is precisely the low-energy limit of Type II string theory compactified on Calabi-Yau 3-folds, where 
one has the microscopic string counting formulas. Thus, by using our results, one can join the dots cleanly between the microscopic and macroscopic pictures of black holes in the same string compactification to~$\IR^4$ for any value of the moduli.

The theory that we discuss is exactly the setup of the original attractor mechanism. The solution that we discuss carries the same charges as the extremal attractor black hole, but 
has the topology of a smooth-capped cigar (times an~S$^2$) rather than an infinite cylinder. 
Equivalently, the solution can be presented as a generalization of the Israel-Wilson-Perj\'es (IWP) solution~\cite{Perjes:1971gv,Israel:1972vx} to a theory with scalar fields.\footnote{In fact, our geometries reduce to the IWP solution when the moduli are fixed to their attractor value everywhere, an implicit assumption in \cite{Iliesiu:2022kny} and \cite{H:2023qko} which we can now justify.}
In this presentation,  the total charge of the black hole gets divided into two harmonic 
sources---corresponding to the north and south poles of the rotating black hole. 
In addition to the electromagnetic fields generated by the original black hole charges, there are new dipole fields in the solution.  
The condition of preserving half the supersymmetry~\cite{LopesCardoso:2000fp} combined with the smoothness of the solutions
fixes all the parameters of the solution in terms of the total charges of the black hole.

\smallskip

Remarkably---although there is no AdS$_2$ near-horizon geometry---all the parameters of the solution are fixed by a set of stabilization equations, which have the same algebraic form as the old extremal attractor equations.
In particular, the imaginary parts of the fixed scalars are given by the total magnetic charges, while the 
corresponding real parts are determined by the electromagnetic charges through algebraic equations 
governed by the prepotential of supergravity. 
However, the spacetime interpretation of the solutions is now different: the geometry is capped, 
the values of the scalars are fixed to their attractor values at the north and south poles of the rotating black hole, and the real part of the scalars now corresponds precisely to the 
dipole charges! 
The on-shell action of the solution is the thermodynamic grand-canonical free energy in accord with the Gibbons-Hawking action principle~\cite{Gibbons:1976ue}. 
Further, although the angular momentum and the area of the horizon are non-trivial functions of the temperature and asymptotic moduli, 
the free energy is independent of the asymptotic moduli and has a trivial temperature dependence exactly as in the last equation of~\eqref{eq:index}. This is the new attractor mechanism for the index.  

\smallskip

The remainder of this paper is structured as follows. In Section~\ref{sec:simplest-saddle-grav-index} we provide a pedagogical example of the simplest saddle-point for an index by studying the truncation of pure supergravity to Einstein-Maxwell theory.
In Section~\ref{sec:sugrarev} we present the technical aspects of ungauged $\mathcal N=2$ supergravity that will be necessary to find the new attractor saddles and review the standard attractor mechanism for extremal black holes. 
In Section~\ref{sec:new-attractor} we find the new Euclidean saddle that contributes to the supersymmetric index and introduce the new attractor mechanism that this solution exhibits. 
We find the on-shell action of this solution in Section~\ref{sec:on-shell-action-and-free-energy-of-the-index} and, as a consequence of the attractor mechanism, we show that the on-shell action is fixed to a value that is independent of the asymptotic moduli. 
We summarize our results and discuss future directions in Section~\ref{sec:discussion}.

\smallskip

\section{The simplest saddle for the gravitational index}
\label{sec:simplest-saddle-grav-index}

We begin by considering the simple set-up of Einstein-Maxwell theory 
\be \label{MaxEins}
S\= \frac{m_\text{P}^2}{16 \pi } \int \, \dd^4x \, \sqrt{-g} \, \bigl(R - F^2 \bigr) \,.
\ee
In this section we work in units such that $m_\text{P}^2=1$. This action is also the bosonic action of pure ungauged supergravity. In this context we discuss solutions of~\eqref{MaxEins} that contribute as saddle-point configurations to the 
gravitational path integral that computes the index~\eqref{eq:index}. These are also solutions of $\mathcal{N}=2$ supergravity coupled only to hypermultiplets\footnote{We will see in Section~\ref{sec:some-examples} that the solutions of this section also work in the presence of vector multiplet, but only for special values of the moduli.}. 
As we see in the following sections, the saddle-point solution discussed in this section 
generalizes to supersymmetric solutions of the full supergravity 
including vector multiplets, when the boundary conditions for the scalar fields are tuned appropriately.

Each saddle point contribution to a gravitational index comes from a spacetime geometry that preserves 
supersymmetry, i.e.~it admits globally well-defined Killing spinors.\footnote{Solutions that do 
not have globally well-defined Killing spinors but that satisfy the index boundary conditions 
give rise to additional zero-modes of the gravitino, corresponding to broken supersymmetry, 
in the full theory.} 
Consider solutions with total electric charge~$Q$ as measured from asymptotic infinity. 
The simplest spherically-symmetric black hole geometry that has globally well-defined Killing spinors is the 
extremal Reissner-Nordstrom solution. In Lorentzian signature, the metric is characterized 
in terms of a single harmonic function $V(\xvec)$,
\be \label{eq:extremalMetric}
\dd s^2 \= -\frac{1}{V^2} \dd t^2 + V^2 \dd x^i \dd x^i \,, \qquad V(\mathbf x) \= 1 + \frac{Q}{|\mx-\mx_{BH}|}\,.
\ee
Here $\xvec=(x_1,x_2,x_3)$ is a coordinate system on the three-dimensional flat base space,
and $\mx_{BH}$ is the location of the black hole on the base space. 
The non-zero components of the field strength are given by~$F_{0i} = \partial_i V^{-1}$ 
with~$\p_i=\p/\p x^i$, $i=1,2,3$.
The Bekenstein-Hawking entropy of this extremal black hole, as given by a quarter of its horizon area 
is~$S^\text{extremal}(Q) = \pi Q^2$. This solution can be generalized to a multi-center extremal black holes system by adding more poles in $V$ at the location of the horizons, as found by Majumdar and Papapetrou \cite{Majumdar:1947eu,Papaetrou:1947ib}.  

When analytically continuing this metric to Euclidean signature, because of the double pole in the emblackening factor, $V^{-2}$, Euclidean time can only be identified periodically with an infinite period, which is incompatible with the boundary conditions required in the computation of the gravitational index \eqref{eq:index} for arbitrary $\beta$.\footnote{
\label{footnote:making-extremal-BHs-periodic}As mentioned in the introduction, an alternate possibility is that Euclidean time can be periodically identified to have period $\beta$ at the expense of creating a cusp that is at an infinite proper distance from any point with $\mathbf x \neq \mathbf x_{BH}$. 
This is the approach taken in \cite{Hawking:1994ii}, which showed that the on-shell action of such a configuration is proportional to~$\beta$, thus naively suggesting that the entropy of such extremal black holes vanishes. 
Moreover, the one-loop determinant around such a configuration in supergravity theories is exactly vanishing \cite{Iliesiu:2021are}, suggesting that such extremal geometries with Euclidean time periodically identified yield no contribution to the supergravity path integral with periodic boundary conditions. }

\bigskip

The metric \eqref{eq:extremalMetric}, however, is not the unique geometry that locally solves the Einstein-Maxwell equations and the Killing spinor equation. The more general solution, 
which we shall express directly in Euclidean signature, can be expressed in terms of two possibly different harmonic functions, $V$ and $\widetilde V$ ($\nabla^2 V = \nabla^2 \widetilde{V} = 0$) \cite{Tod:1983pm},
\be \label{eq:IWPmetric}
 \dd s^2 \= \frac{1}{V \widetilde{V}} (\dd t_E + \omE)^2 + 
V \widetilde{V} \dd x^i \dd x^i \,.
\ee
Such a solution carries a non-trivial angular velocity that can be characterized in terms of a three-dimensional 
gauge field $\omE$ on the base space $\mathbf x$ that satisfies 
\be \label{eq:omega_equation}
\nabla \cross \omE \= \widetilde{V} \nabla V - V \nabla \widetilde{V} \,.
\ee 
The field strength in this case, is given by, 
\begin{align}\label{fsiw}
F_{0i} &= -\frac{\i }2 \partial_i\left(\frac{1}{V}+ \frac{1}{\widetilde V}\right)\,, \qquad F^{ij} = \frac{-\i}{2\sqrt{g}} \, \varepsilon^{ijk} \partial_k \left(\frac{1}{V}-\frac{1}{\widetilde V}\right)\,.
\end{align}
The above solution was discovered by  Perj\'es, Israel, and Wilson \cite{Perjes:1971gv, Israel:1972vx}. Its analytic continuation to Euclidean signature has been extensively studied in \cite{Whitt:1984wk, Yuille:1987vw} (see also \cite{Dunajski:2006vs}).  The simplest such solution, which has total charge $Q$, has
\be 
\label{eq:choice-of-V-tilde-V}
V \= 1+ \frac{Q}{|\xvec - \xvec_N|} \,, \qquad \widetilde{V} \=1+ \frac{Q}{|\xvec - {\xvec}_S|} \,.
\ee
for two points in base space $\xvec_N$ and $\xvec_S$. This solution is asymptotically flat, and 
when $V \neq \widetilde V$ (or $\xvec_N\neq \xvec_S$),  it is non-extremal and
has a non-trivial profile for $\omE$ as seen from~\eqref{eq:omega_equation}.
The solution has an electric field, whose electric flux around any smooth closed surface enclosing~$\xvec_N$ 
and~${\xvec}_S$ determines the charge of the black hole, as well as a magnetic field, whose magnetic flux 
around any smooth closed surface vanishes. Far away from either  $\mathbf x_{N}$ or $\mathbf x_{S}$, 
the electric field looks like that of a point charge (decaying like~$1/r^2$), while the magnetic field 
behaves like that of a magnetic dipole (decaying like~$1/r^3$).   
Along the axis between $\mathbf x_{N}$ and  $\mathbf x_{S}$ the magnetic field points away 
from~$\mathbf x_{N}$, which can therefore be identified as the north pole of the solution, 
and towards~$\mathbf x_{S}$, which is therefore identified as the south pole of the 
solution.\footnote{Here, we consider the case~$Q>0$. For $Q<0$, the two poles are switched.} Since $V$ and $\widetilde{V}$ only need to be harmonic, there is a straightforward generalization where $V$ and $\widetilde{V}$ have multiple centers.

In the following presentation we perform, for simplicity,  a diffeomorphism 
such that $\xvec_N = (0,0,\alpha)$ and $\xvec_S = (0,0,-\alpha)$, and we denote by $\phi$ the $2\pi$-periodic angular coordinate around the $z$-axis 
passing through the north and south pole.

The above solution is actually a Euclidean black hole solution. 
We make this more explicit below using a mapping to a more familiar set of coordinates, 
but before doing so, we extract various properties of this solution in the above coordinates. 
These manipulations turn out to very useful, especially in the presence of vector multiplets which we analyze in the rest of this paper, and also in more general Euclidean configurations 
that we discuss in a future publication \cite{wip}.

\begin{figure}
    \begin{center}
    \begin{tikzpicture}
\draw [blue!60, thick] (0,-2) -- (0,2);
\draw [gray] (0,0) ellipse (1 and 2);
\draw [gray] (0,0) ellipse (1 and 0.1);
\filldraw [black] (0,2) circle (0.08);
\filldraw [black] (0,-2) circle (0.08);
\node at (0.4,-2.2) {\footnotesize $\mathbf x_{S}$};
\node at (0.4,2.2) {\footnotesize $\mathbf x_{N}$};
\node at (1.2,0) {\footnotesize $\Sigma$};
\node at (0,-3) {\footnotesize $(a)$};
    \end{tikzpicture}
    \hspace{2cm}
        \begin{tikzpicture}
\draw [blue!60, thick] (0,-2) -- (0,2);
\draw [gray] (0,2) ellipse (1 and 1);
\draw [gray] (0,2) ellipse (1 and 0.1);
\filldraw [black] (0,2) circle (0.08);
\filldraw [black] (0,-2) circle (0.08);
\draw [gray] (0,1.2) ellipse  (0.6 and 0.05);
\draw [gray] (0,2.8) ellipse  (0.6 and 0.05);
\node at (1,1.2) {\footnotesize $ \mathcal{C}_1$};
\node at (1,2.8) {\footnotesize $\mathcal{C}_2$};
\node at (0.4,-2.2) {\footnotesize $\mathbf x_{S}$};
\node at (0.4,2.2) {\footnotesize $\mathbf x_{N}$};
\node at (0,-3) {\footnotesize $(b)$};
    \end{tikzpicture}
    \hspace{2cm}
           \begin{tikzpicture}
\draw [thick] (0,2) ellipse  (1 and 0.1);
\draw [thick] (-1,2) arc (180:360:1 and 3.7);
\draw [ thick] (1.7,2) circle (0.6);
\draw [fill = blue!20, thick] (1.7,-1.5) circle (0.4);
\draw [thick] (1.1,2) to (1.3,-1.5);
\draw [thick] (2.3,2) to (2.1,-1.5);
\draw [thick, ->] (1.4,2) to [bend right = 20] (2,2);
\draw [thick, ->] (1.4,-1.5) to [bend right = 20] (2,-1.5);
\filldraw [black] (1.7,-1.1) circle (0.07);
\filldraw [black] (1.7,-1.9) circle (0.07);
\node at (2.5,-1.1) {\footnotesize $\mathbf x_{N}$};
\node at (2.5,-1.9) {\footnotesize $\mathbf x_{S}$};
\node at (0,-3) {\footnotesize $(c)$};
    \end{tikzpicture}
    \end{center}
    \caption{\footnotesize (a) The figure shows a spatial slice of the geometry and indicates the poles at $\mathbf x_N$ and $\mathbf x_S$ joined by the Dirac string (blue line), surrounded by the surface $\Sigma$. If we were in flat space, as the width of $\Sigma$ goes to zero, the area would also vanish. In the geometry specified by \eqref{eq:IWPmetric}, as the width of $\Sigma$ vanishes the area remains finite and non-zero. This line is the location of the black hole horizon, as explained in the text. (b) Geometry of the volume (the sphere around $\mathbf x_{N}$) used to derive \eqref{omOmrel0}. The circles $\mathcal{C}_{1,2}$ have an infinitesimal radius, so the volume integral over the sphere and between $\mathcal{C}_1$ and $\mathcal{C}_2$ are identical. (c) Diagram of the Euclidean section of the rotating black hole with topology $D^2 \times S^2$. There is a time-like circle that contracts at the horizon. The horizon is represented by the blue line in (a) and (b), and by the blue sphere in (c). }
    \label{fig:enter-label}
\end{figure}
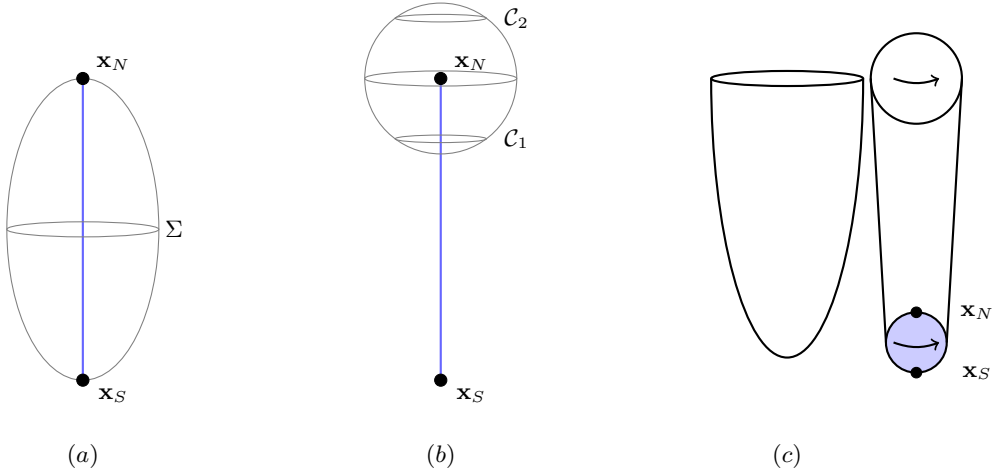

Firstly, we identify the horizon of this Euclidean solution \cite{Israel:1972vx, Hartle:1972ya}. Consider a closed surface that is infinitesimally close in base space to the line 
joining $\mathbf x_{N}$ and $\mathbf x_{S}$. Although it looks like a degenerate surface
in the flat coordinates, the proper area of this surface is non-zero, and therefore
the line connecting $\mathbf x_{N}$ and $\mathbf x_{S}$ is actually a bubble. 
In fact, this surface minimizes the area of surfaces joining the two points, and so 
it should be identified with the horizon of the Euclidean black hole geometry. 
Indeed, the value of~$\omE = \omEphi \, \dd \phi$ is constant along the line 
joining~$\mathbf x_{N}$ and~$\mathbf x_{S}$, as consistent with the interpretation 
as a horizon. 
This can be seen by calculating the divergence of both sides of the 
equation \eqref{eq:omega_equation} and then integrating over a closed volume 
that intersects this line and contains $\mathbf x_{N}$~\cite{Yuille:1987vw}. 
Using Stokes theorem twice to relate the integral over this volume to an integral 
over a surface and then to an integral over a closed loop we obtain  
\be \label{omOmrel0}
2\pi \omEphi  \= \int_{\partial S \rightarrow 0} \omega_E \= 4\pi Q \, \widetilde{V}(\xvec_N) \,.
\ee
Here $\partial S$ denotes an infinitesimal loop placed anywhere along the line 
from $\mathbf x_{N}$ to $\mathbf x_{S}$ and it is clear from the right-hand side that 
the value of~$\omEphi$ is a constant on the horizon. 
Upon evaluating the expressions on both sides of~\eqref{omOmrel0}, we obtain
\be \label{omOmrel}
2\pi \omEphi|_{\text{hor}} \= 4\pi Q\left(1+\frac{Q}{2\alpha}\right) \,.
\ee
The area of the horizon is given by 
\be  \label{eq:area-bubble}
A_\text{H} \= 2\pi \,   \abs{\xvec_N - {\xvec}_S} \left(\omega_{E \,\phi}|_{\text{hor}} \right)
 \= 4\pi  \a \left(\omEphi|_{\text{hor}} \right) \,.
\ee

The constant value~$\omEphi|_\text{hor}$ is simply related to the angular velocity of the black hole. 
To see this, consider the coordinate transformation
\be \label{eq:t-coord-transform-for-smoothness}
t_E \; \to \; t_E' \= t_E +  \frac{1}{\Omega_E} \phi ,
\ee
which induces the gauge transformation 
\be \label{eq:t-coord-transform-for-smoothness2}
\omE \;\to \;
\omE' \= \omE - \frac{1}{\Omega_E} \dd \phi\,.
\ee
Upon setting~$\omE' =0$, one obtains $\Omega_E = 1/(\omEphi|_{\text{hor}})$. 
In the Lorentzian theory this corresponds to a coordinate system in which the horizon does not rotate,
that is to say that~$\Omega_E$ is the Euclidean angular velocity of the black hole, 
related to the Lorentzian angular velocity $\Omega$ via Wick rotation, $\Omega = \i \Omega_E$. 

Note that we could have equally well enclosed the south pole by a closed surface 
that is punctured by the line between $\mathbf x_{N}$ and $\mathbf x_{S}$, and the 
symmetry of the expression~\eqref{eq:area-bubble} makes it clear that the 
above steps would have yielded the same equation. 
This shows that the spacetime can only be smooth if the residues of~$V$ at~$\mathbf x_{N}$ 
and of~$\wt V$ at $\mathbf x_{S}$ are equal. 
This is why we set the values of the residue in \eqref{eq:choice-of-V-tilde-V} to be equal. 

\medskip 

We still need to ensure that the entire manifold is smooth when the Euclidean 
time~$t_E$ is compact with period~$\beta$. 
The fact that the integral of $\omE$ over a circle with infinitesimal proper length does not 
vanish implies that there is a Dirac string for the gauge field $\omE$ between $\mathbf x_{N}$ 
and $ \mathbf x_{S}$. To improve on the dipole picture that we have hinted at earlier, because of 
the Dirac string in $\omE$, the two poles can be seen as a Taub-NUT and anti-Taub-NUT pair\footnote{The reason for this terminology is the following. When one of the two harmonic functions, which for concreteness we take here to be $\tilde{V}$, is constant throughout space, the solution \eqref{eq:IWPmetric} becomes the Taub-NUT geometry. This might be unexpected since the Taub-NUT geometry is a vacuum solution. The reason is that the field strength $F_{\mu\nu}$ in \eqref{fsiw} becomes selfdual, when either $V$ or $\tilde{V}$ is constant, producing a vanishing stress tensor. Given this, as long as the poles of $V$ and $\tilde{V}$ do not coincide, the solution close to any pole will look like Taub-NUT. }. For the geometry to be smooth, we impose that this is only a coordinate singularity and can be 
removed by a coordinate transformation. 
For this we need to ensure that the transformation~\eqref{eq:t-coord-transform-for-smoothness} which sets~$\omega_E' = 0$
is consistent with the periodicities of the $\phi$ as well as of the $t_E$ coordinate. 
Since for any $t_E$, $(t_E, \phi) \sim (t_E, \phi+2\pi)$, the diffeomorphism \eqref{eq:t-coord-transform-for-smoothness} only results in a valid gauge transformation \eqref{eq:t-coord-transform-for-smoothness2}
if \cite{Hartle:1972ya}
\be \label{Ombetarel}
\Omega_E \= -\i \Omega \= \frac{2\pi}\beta.
\ee
Finally, the equations \eqref{omOmrel} and~\eqref{Ombetarel} together determine the base-space distance 
between the north and south pole to be 
\be 
\label{eq:alpha-from-smoothness}
\alpha \= \frac{2\pi Q^2}{\beta-4\pi Q}\,.
\ee

\medskip

To summarize, by separating the poles in the extremal solution \eqref{eq:IWPmetric}, 
we have found a rotating solution whose boundary conditions are those necessary in the saddle-point computation of an index. 
Smoothness completely fixes all parameters of this solution, i.e.~the angular velocity of the black hole~$\Omega$ 
and the distance in base-space between~$\mathbf x_{N}$ and~$\mathbf x_{S}$, in terms of $\beta$ and $Q$.

\medskip

The expression~\eqref{Ombetarel} is precisely the boundary condition that leads to the insertion~$e^{2\pi \i J}$ in the index~\eqref{eq:index}. Phrased differently, the most general Euclidean solution in Einstein-Maxwell theory in asymptotically flat space 
that supports Killing spinors is only smooth for boundary conditions corresponding to the gravitational index. It is important to stress that the converse of this statement is not true; the boundary conditions in the gravitational path integral are consistent with supersymmetry as long as $\beta \Omega_E = 2\pi (1+2n)$ for any integer $n$. Nevertheless, the solution does not preserve supersymmetry in the bulk unless $n=0$ or $-1$ ($n$ and $-1-n$ are related by a rotation) as explained in \cite{Iliesiu:2021are}. Finally, note that the Wick rotation of the extremal metric \eqref{eq:extremalMetric}, which also admits  
globally defined Killing spinors is only consistent with the limit $\beta \to \infty$, $\Omega_E\to 0$.

Recall that a reason it was previously thought that the index is not directly captured by a saddle associated with a 
Euclidean black hole solution comes from considering  the periodicity condition for fermionic fields at the asymptotic boundary.\footnote{For instance, see \cite{Witten:1998zw} for an early discussion in the context of AdS/CFT.} 
On one hand, the insertion of $(-1)^{\sf F}$ implies that that spinors should be periodic around the thermal circle. On the other hand, the time circle is typically contractible and 
smoothness therefore only admits anti-periodic spinors. 
Thus it would seem that there is no well--defined spin-structure in the full geometry. 
This is not the case for the solution~\eqref{eq:IWPmetric}. 
Due to the rotation of the black hole, the true contractible circle is a combination of time and one of the angles,  
and is generated by~$\partial_{t_E} + \frac{\beta}{2\pi}\partial_{\phi}$. Thus, all fermionic fields satisfy, 
\be
\psi(t_E, \phi) \= -\psi(t_E+\beta, \phi+2\pi)\,, \qquad \psi(t_E, \phi) \= \psi(t_E+\beta, \phi)\,. 
\ee
Thus we obtain a well-defined smooth spin-structure that is 
consistent with the fermionic boundary conditions for the index. 
For this idea to work it is important to have the fermions to be charged under a bosonic R-symmetry gauge field which participates in the 
solution~\cite{Cabo-Bizet:2018ehj, Cabo-Bizet:2019osg}.\footnote{This could be a vector gauge field as in gauged supergravity in AdS space or spacetime angular momentum in ungauged supergravity in flat space~\cite{Iliesiu:2021are}.}

\medskip

Before we proceed to evaluate the Einstein-Maxwell action at the saddle point, it is useful to perform a coordinate change that puts the metric \eqref{eq:IWPmetric} in a more recognizable form. Defining, the new coordinates  $r$, $\theta$, and $\phi$ through \cite{Whitt:1984wk}, 
\begin{align}
 a 
 \,\equiv 
 \, \i\alpha\,, \qquad x +\i y \, \equiv \,  \sqrt{(r-Q)^2 + a^2} \, \sin \theta \, \rme^{\i \phi} \,, \qquad z \,\equiv \, (r-Q) \cos \theta \,.
 \label{eq:coordinate-change}
\end{align} 
The metric \eqref{eq:IWPmetric} consequently becomes 
\be \label{KNmetric}
\begin{split}
\dd s^2 & \=  \frac{\Delta}{\rho^2} (\dd t_E + \i a \sin^2 \theta \dd \phi )^2 + \frac{\rho^2 \dd r^2}{\Delta} + \rho^2 \dd \theta^2 +
\frac{\sin^2 \theta}{\rho^2} (a \dd t + (r^2 + a^2) \dd \phi )^2 \,,\\
\rho &\= \sqrt{r^2 + a^2 \cos^2 \theta } , \qquad 
\Delta \= (r-Q)^2 + a^2 \,, \qquad  a\= \i \alpha = \i \frac{2\pi Q^2}{\beta-4\pi Q} \,.
\end{split}
\ee
This is nothing but the Euclidean Kerr-Newman metric in Boyer-Lindquist coordinates with charge $Q$, mass $M=Q$, temperature $\beta^{-1}$, and angular velocity $\Omega_E = 2\pi/\beta$. This is precisely the solution found in \cite{Iliesiu:2021are} as the leading saddle to the gravitational index in a supergravity theory truncated to Einstein-Maxwell \eqref{MaxEins}. 
It is the flat space analog of the earlier found asymptotically AdS$_5$ solution \cite{Cabo-Bizet:2018ehj}, which represented the leading gravitational saddle-point contribution in the dual of the $\mathcal N=4$ SYM refined index. 
In Lorentzian signature, the roots~$r_{\pm}$ of~$\Delta(r)=0$ define the location of the outer and inner horizons, with the outer horizon $r_+$ becoming our Euclidean horizon where the spacetime smoothly pinches off. As advertised, \eqref{eq:coordinate-change} and the expression of $\Delta(r)$ in \eqref{KNmetric} show that the horizon at $r=r_+$ is mapped by our coordinate transformation to the line at $x=y=0$ with $z\in[-a, a]$, along which we identified our Dirac string. 

\medskip

The Euclidean on-shell action of the above configuration can be computed in in a straightforward manner in 
either coordinate system and one obtains\footnote{The computation is simplified by the fact that the Ricci scalar $R = 0$ 
on the entire spacetime and that the Maxwell term can be written as a boundary term proportional to the 
electron-magnetic boundary term that is necessary when fixing the field strength at the asymptotic boundary. 
While we shall not review the derivation of the on-shell action here, a similar derivation will be performed 
in Section~\ref{sec:new-attractor} for the more general supergravity coupled to vector multiplets in 
which scalar fields vary throughout the geometry.}
\be
\label{eq:on-shell-action}
-I_E^\text{on-shell} \= - \beta Q + \pi Q^2\,.
\ee
We recognize the first term $-\beta Q$ as arising from the BPS energy of the black hole and the second term $\pi Q^2$ as the entropy of the extremal black hole~\eqref{eq:extremalMetric}.
\footnote{The term that is linear in $\beta$ can also be eliminated by adding a boundary counter-term 
that shifts the ground state energy to zero in a fixed charge sector.}

Note that this extremal entropy is not the horizon area of the Euclidean black hole at finite temperature~$\beta$. 
Indeed, upon imposing the smoothness relations in \eqref{eq:area-bubble}, we find 
\be 
\label{eq:area-horizon-KN}
\frac{A_H}4 \= 2\beta \alpha 
\= \frac{\pi \beta Q^2}{\beta - 4\pi Q} \,,
\ee
which depends on~$\beta$. The solution works in a straightforward way as long as $\beta> 4\pi Q$. The area diverges at $\beta = 4\pi Q$ and becomes negative at $\beta<4\pi Q$. This leaves two options. First, the solution ceases to contribute to the path integral at high temperatures $\beta < 4\pi Q$. This would imply a discontinuity of the index as a function of temperature, reminiscent of the wall-crossing phenomena in terms of the moduli variation. If such a jump takes place at $\beta = 4\pi Q$ this would be a wall-crossing phenomena of a totally new type\footnote{Depending on how the index is defined, it can depend on the temperature for quantum theories with a continuum spectrum \cite{Imbimbo:1983dg}.}. Second, we keep the $\beta < 4\pi Q$ solution as a complex saddle such that the index remains temperature independent. A better understanding of the strongly coupled quantum mechanical dual of this black hole would elucidate which option is correct, but for now we focus on $\beta > 4\pi Q$ and continue with our analysis. The horizon area is not the canonical transform of the on-shell action~\eqref{eq:area-horizon-KN}
with respect to the temperature,  
\be 
\frac{A_H}4 \; \neq \; -\left(1- \beta \partial_\beta \right)I_E^\text{on-shell} \,. 
\ee
Such a relation would only be valid when the thermodynamic potentials are independent of each other. The index calculation is defined at fixed temperature~$\beta$ \emph{and} 
fixed chemical potential for angular momentum, satisfying a constraint between them. The on-shell action~\eqref{eq:on-shell-action} 
should, therefore, be interpreted as the free energy associated with the corresponding grand-canonical ensemble 
as obtained from the Gibbons-Hawking quantum statistical relation \cite{Gibbons:1976ue}. 
To verify this, one can read off the ADM mass~$M$ of the solution and the Euclidean angular 
momentum~$J_E = -\i J$ from the fall-off of \eqref{eq:IWPmetric} or \eqref{KNmetric} as $r\to \infty$, 
\be \label{eq:angular-momentum-KN}
    M\=Q \,,   \qquad J_E \=  Q \alpha \= \frac{2\pi Q^3}{\beta-4\pi Q}  \,.
\ee 
Using these values of the area and the charges and the smoothness condition \eqref{eq:alpha-from-smoothness}, it is then easy to check that the on-shell action satisfies
\be 
\label{eq:free-energy-IW}
-I_E^\text{on-shell} \; \equiv \; - \beta F 
\= - \beta M - \underbrace{\beta \Omega_E}_{=2\pi} J_E + \frac{A_\text{H}}{4} \,.
\ee

To summarize,  the saddle-point of the index is a Euclidean spinning black hole with an electric charge.
The solution has a magnetic dipole field which is completely determined in terms of the monopole 
electric charge. 
The area of the black hole, its angular velocity, and its angular momentum all have non-trivial 
temperature dependence, but its free energy only has trivial temperature dependence through 
the~$M_\text{BPS}$ term, and the corresponding entropy is independent of the size of the thermal circle at infinity.

This can be viewed as the simplest version of the new attractor mechanism in the Euclidean computation 
of the supersymmetric index. 
In the following sections, we generalize our analysis to compute the gravitational index in a 
supergravity theory with an arbitrary number of vector multiplets. Aside from serving as a pedagogical example, the solution presented in this section also reappears per se 
in the more complicated examples in Section~\ref{sec:some-examples} upon fixing the scalar fields in supergravity to their attractor values.

\section{A review of $\CN=2$ supergravity and the classic attractor mechanism \label{sec:sugrarev}}

In this section we review the classic attractor mechanism for supersymmetric 
black holes in four-dimensional asymptotically flat space~\cite{Ferrara:1995ih, Ferrara:1996dd}. 
The attractor mechanism, as initially formulated, applies 
to $\frac12$-BPS black holes in theories of~$\CN=2$ supergravity 
(8 supercharges) coupled to a number of vector multiplets. 
The basic phenomenon is that near the horizon of a BPS black hole the shape 
of the metric, gauge fields, and the scalar fields all get fixed by the charges of the black hole. 
At the fixed point, the metric has the form of~AdS$_2 \times$S$^2$, and 
the electromagnetic field strengths and the scalar fields are constant.
The near-horizon region is actually a fully supersymmetric solution 
in its own right. 

In the discussion below of the attractor mechanism in~$\CN=2$ supergravity, 
we follow the presentation of the reviews~\cite{Mohaupt:2000mj, Pioline:2006ni}. 
The full geometry of the $\frac12$-BPS black hole can be derived as a 
consequence of supersymmetry combined with (often implicit) assumptions 
about smoothness of the Lorentzian geometry (see 
e.g.~\cite{Mohaupt:2000mj}).\footnote{The fact that the near-horizon geometry is fixed in terms of 
the charges can be explained---without using supersymmetry---as a consequence 
of the extremal nature of the Loretzian BPS black hole. 
This is summarized by the elegant entropy function formalism~\cite{Sen:2005wa}.
Here our eventual goal is to \emph{remove} the extremality condition, and so 
we follow the route of supersymmetry. }
The BPS equations 
are recast as a set of first-order equations for the metric, gauge, 
and scalar fields. The near-horizon configuration is a fixed point of 
this first-order syste---hence the name ``attractor"---which preserves full  
supersymmetry.
In particular, the scalar fields can start at an arbitrary value (within 
a basin of attraction) at asymptotic infinity and the first-order system 
describes a flow in the space of scalar field values ending at the attractor
value at the horizon.  

The attractor solutions can be lifted to more general theories of supergravity 
with other~$\CN=2$ multiplets like hypermultiplets, as well as to theories 
with extended supersymmetry. In such situations, the other fields typically 
do not participate in the solutions, and so we do not discuss them here. 
In the context of string theory, the attractor black holes are realized in 
Calabi-Yau compactifications and, in fact, there is a very beautiful 
interpretation of the attractor flow equations in the Calabi-Yau moduli 
space~\cite{Moore:1998pn,Denef:2007vg}. Although we do not discuss this in detail, 
it useful to use the language of the Calabi-Yau compactification and 
we present many of the formulas in that notation. 

In Section~\ref{sec:sugrabackgnd} we review some basic aspects of~$\CN=2$ supergravity coupled to 
vector multiplets as needed for our presentation. 
In Section~\ref{subsec:BPSeqns} we review the most general~$\frac12$-BPS solutions of this theory.
In Section~\ref{sec:old-attractor-black-hole} we review the classic attractor mechanism.
In Section~\ref{sec:spacetime-dep} we review the spacetime dependence of the general BPS solutions.

\subsection{$\CN=2$ supergravity coupled to vector multiplets \label{sec:sugrabackgnd}}

We consider four-dimensional~$\CN=2$ supergravity (8 real supercharges) coupled 
to vector multiplets in the superconformal formalism~\cite{deWit:1979dzm,deWit:1980lyi}.   This is an elegant formalism in which supersymmetry is realized off-shell, 
and the symplectic invariance of the theory manifest. The local gauge symmetry is enlarged from a local super-Poincar\'e symmetry to 
a local superconformal symmetry. The field content consists of one Weyl multiplet, $\nv+1$ vector multiplets 
labeled by $I=0,1,\dots, \nv$, 
and some compensator multiplets whose role is to gauge-fix some of the 
extra gauge symmetries (see~\cite{Mohaupt:2000mj} for a nice review). 

The Weyl multiplet contains the vielbein and its superpartners, and 
each vector multiplet contains a vector field~$A^I$, a complex scalar~$X^I$,
as well as gaugini and auxiliary fields. 
The geometry of the moduli space of the scalar fields in the vector multiplets, 
called the vector multiplet moduli space, plays an important role in the physics. 
In particular, it  governs the kinetic terms and the couplings of the above 
supergravity theory. 
This moduli space is a projective special K\"ahler manifold of complex dimension~$\nv$,
for which the scalars~$X^I$, $I=0,\dots,\nv$ are projective (or homogeneous) 
coordinates. The scaling~$X^I \to \lambda X^I$ is part of the gauge symmetry 
of the superconformal description, called the Weyl symmetry, under which~$X^I$ 
has weight one. 
In order to reach Poincar\'e supergravity, one fixes the Weyl symmetry by choosing a field with non-zero Weyl weight and setting it equal to the 
physical scale~$m_\text{P}$.\footnote{Similarly, additional gauge symmetries of the superconformal theory like the~$U(1)_R$ symmetry are fixed by giving expectation values to fields charged under those symmetries.}

A central role in the formalism is played by the electric-magnetic 
duality of~$\CN=2$ supergravity. The equations of motion of the two-derivative 
supergravity theory transform linearly under~$Sp(2\nv+2;\IR)$ transformations
which is the electric-magnetic duality group of~$\CN=2$ supergravity, under which the magnetic and electric charges~$(P^I,Q_I)$ transform as a vector.  
One then completes the~$\nv+1$ scalars~$X^I$ to a~$2(\nv+1)$-dimensional vector~$(X^I,F_I)$ under~$Sp(2\nv+2)$. 
At generic points in the moduli space, $F_I$ can be expressed in terms of the \emph{prepotential}~$F(X)$
and vice-versa, as 
\be \label{FFIrel}
F_I \= \frac{\p}{\p X^I} F(X)  \,, \qquad F \= \frac12 \, X^I F_I \,.
\ee
The prepotential is a homogeneous function of degree (and therefore Weyl weight) two, 
which is reflected in the second equation. 
The prepotential function gives a very convenient description of the physics---indeed, the 
two-derivative action is completely determined by~$F$, and hence all physical quantities 
are computable as a function of~$X$ and~$\overline{X}$.\footnote{It is, however, important to note 
that the prepotential breaks manifest~$Sp(2\nv+2)$ duality invariance, 
it is not unique (different choices of~$F$ may give the same equations of motion modulo duality rotations), 
and it may not exist at all points in the moduli space.}

At a formal level, one introduces a principal~$Sp(2\nv+2)$-bundle and the associated 
holomorphic vector bundle~$E_V$, and~$(X^I,F_I)$ are the coordinates of a section~$\Omega_{\text{hol}}$ of~$E_V$~\cite{Pioline:2006ni}.
In the context of Calabi-Yau compactification of Type II string theory, 
the projection of $\Omega_{\text{hol}}$ at a given point in moduli space 
is identified with an element of cohomology (middle-dimensional cohomology for Type IIB 
and even cohomology for Type IIA), and is called the~\emph{period vector}. 
This identification defines a natural intersection product on these spaces 
given by integrals over the Calabi-Yau manifold.

We denote symplectic vectors as~$A=(\wt{A}^I,A_I)$ with~$I=0,1,\dots, \nv$ and their components also as~$A^\alpha$, $\alpha=1,\dots, 2\nv+2$. 
The symplectic product of two vectors~$A=(\wt{A}^I,A_I)$ 
and~$B =(\wt{B}^I, B_I)$ is given by
\be
\langle A ,  B \rangle 
\= \wt A^I  B_I - \wt B^I A_I 
\; \equiv \; I_{\alpha \beta}
A^{\alpha} B^\beta \,,
\ee
where $\a,\,\beta=1,\,\dots,2n_V+2$.
In the context of the Calabi-Yau compactification, the matrix~$I_{\alpha\beta}$ is the 
intersection product matrix in the symplectic basis.

We denote the holomorphic period vector~$\Omega_\text{hol}$ 
and the charge vector~$\Gamma$ by  
\be 
\Omega_{\text{hol}} \= (X^I,F_I)\,, \qquad 
\Gamma \= (P^I,Q_I)\,.
\ee
Here $P^I$ and $Q_I$ are the magnetic and electric charges, respectively. The symplectic product~$\langle \, \Gamma \,,\, \wt \Gamma\, \rangle = P^I \wt Q_I - Q_I \wt P^I $
is immediately recognized as the electric-magnetic duality invariant Dirac-Zwanziger product.

We introduce two basic geometric quantities that are important for the 
physics. 
Firstly, we have the generalized K\"ahler potential~$\CK$ defined as 
\be \label{defK}
\rme^{-\CK(X,\overline{X})} 
\= \i \bigl( X^I \overline{F}_I -\overline{X}^I F_I  \bigr) 
\=  \i \intprod{\Omega_{\text{hol}}}{\overline{\Omega}_{\text{hol}}} \,,
\ee
It is clear that~$\rme^{-\CK(X,\overline{X})}$ is a homogeneous function of Weyl weight~2. 
It will also be useful to define the normalized period vector of weight~0 
(which we will simply refer to as the period vector from now on) as
\be 
\Omega \= \rme^{ \frac{1}{2}\CK\left(X, \overline{X}\right)}\, \Omega_\text{hol}\,, \qquad \langle \Omega, \overline{\Omega}\rangle \= -\i.
\ee
Secondly, we have the central charge function of the 
symplectic charge vector 
\be \label{defZ}
Z(\Gamma;\Omega) \= 
\intprod{\Gamma}{\Omega}
\= \rme^{\half\CK(X,\overline{X})}\,(P^I F_I-Q_I X^I ) \,.
\ee
The central charge function has Weyl weight~0, i.e.~it is invariant under scaling of the period vector. The central charge is precisely the central charge of the~$\CN=2$ supersymmetry algebra of the theory,
which enters the BPS bound on the mass in units of~$m_\text{P}$
of any physical configuration 
\be
M^2 \ge |Z(\Gamma,\Omega_\infty)|^2 \, m_\text{P}^2 \,,
\ee
where $\Omega_\infty$ is the period vector evaluated at  asymptotic infinity.

\medskip

\subsection*{The supergravity action} 

The full action of conformal supergravity includes all the compensator multiplets~\cite{Mohaupt:2000mj}
and is invariant under off-shell supersymmetry. 
In this paper we only discuss classical solutions at 2-derivative level, and it is 
useful to directly go to the Poincar\'e supergravity by gauge-fixing the Weyl symmetry
as mentioned above. 
We introduce the physical scale by setting 
\be \label{gaugefixingcondition}
\rme^{-\CK(X,\overline{X})} \= m_\text{P}^2 \,.
\ee
In some of the formulas below, we suppress factors of~$m_\text{P}$.
This gauge-fixing condition has the advantage of preserving the symplectic symmetry.
The~$U(1)_R$ symmetry of the superconformal theory is fixed by e.g.~fixing the phase of~$X^0$ to be zero.  Together, this fixes one combination of the scalar fields, so that one is left 
with $\nv$ propagating complex scalars.

The gauge-invariant ratios~
\be 
t^A\equiv X^A/X^0\,,
\ee 
with $A=1,\dots, \nv$, called the 
special coordinates, are sometimes used to parameterize the 
vector multiplet moduli space. One can invert this to obtain~$X(t)$ up 
to an ambiguity of rescaling by a holomorphic function. 
The metric on moduli space is given by (with~$\p_A \equiv \frac{\p}{\p t^A}$, 
$\p_{\overline A} \equiv \frac{\p}{\p {\overline t}^A}$), 
\be
G_{A \overline B} (t,\overline{t}) \= \p_A \p_{\overline B} \, 
\CK \bigl(X(t),\overline{X}(\overline{t}) \bigr) \,,
\ee
evaluated on the hyperplane defined by~\eqref{gaugefixingcondition}.
It is independent of the holomorphic rescaling and is therefore unambiguous.  

\smallskip

The bosonic action of the super-Poincar\'e theory is 
\be
\frac{1}{m_\text{P}^2} \, S \= \frac{1}{16\pi} \int \dd^4 x \sqrt{-g} R -  \int G_{A\bar{B}} \, \dd t^A \wedge \ast \dd\Bar{t}^{B} 
- \frac{1}{16\pi} \int F^I \wedge G_I \,,
\label{eq:Lorentzian_action}
\ee
where $G_I$ is the dual field strength defined as 
\be \label{FGrel}
G_I \= -\Im \, \CN_{IJ}   \ast F^{J} - \Re \, \CN_{IJ}  F^{J} 
\,,
\ee
so that
\be
-  \int F^I \wedge G_I 
\=  \int \left( 
\Im \, \CN_{IJ} F^{I} \wedge \ast F^{J} + \Re \, \CN_{IJ} F^{I} \wedge F^{J} 
\right) 
\,.
\ee
Here the matrix $(-\CN_{IJ})$, conventionally called the period matrix, 
is given by 
\be
\CN_{IJ}  \= \overline{F}_{IJ} + \i \frac{N_{IK} X^K N_{JL} X^L}{X^M N_{MN} X^N} \,.
\ee 
in terms of the derivatives of the prepotential 
\be
F_{IJ} (X) \= \p_I \p_J F (X) \,, \qquad N_{IJ} \= F_{IJ} - \overline{F}_{IJ}  \,.
\ee

\subsection{General stationary BPS solutions of~$\CN=2$ supergravity \label{subsec:BPSeqns}}

It is a classic result~\cite{Tod:1983pm} that the most general stationary 
solution of pure $\CN=2$ supergravity admitting a Killing spinor has a metric similar to the Israel-Wilson-Perj\'es form~\cite{Israel:1972vx,Perjes:1971gv}, 
\be \label{IWPmetric}
\dd s^2 \= - \rme^{2U} (\dd t+\omega)^2 +  \rme^{-2U} \dd x^m \dd x^m \,,
\ee
where~$U$ is a function of the base space coordinates~$x^m$, $m=1,2,3$, 
and~$\omega=\omega_m \dd x^m$ is a one-form on the base space. In order to have asymptotically flat space, one 
has~$\rme^{-2U(\infty)}=1$. 
A similar result is true in more general supergravities~\cite{Behrndt:1997ny,Kallosh:1994ba}. 
In the present context of~$\CN=2$ supergravity coupled to an arbitrary 
number of vector multiplets, the general $\frac12$-BPS stationary solution is as follows~\cite{Behrndt:1997ny, LopesCardoso:2000qm}.  

The metric takes the form~\eqref{IWPmetric}.\footnote{In the superconformal formalism, the vielbein
has Weyl-weight~$-1$, and so there are appropriate factors of~$m_\text{P}$ in the above 
line element that have been suppressed~\cite{LopesCardoso:2000qm}.}
The  scalar fields obey the so-called \emph{generalized 
stabilization equations}, 
\be \label{genstabOm}
 \i \bigl(\overline{Z}(\Hv;\Omega) \, \Omega - Z(\Hv;\Omega) \, \overline{\Omega} \,\bigr) \= \Hv \,,
\ee
where~$\Hv =(\wt H^I,H_I)$ is a symplectic vector of functions on base space, 
and~$Z$ is the central charge function defined in~\eqref{defZ}.
Here the conjugate function~$\overline{Z}$ is defined as  
\be \label{Zbardef}
\overline{Z}(\Hv;\Omega) \= Z(\Hv;\overline{\Omega}) \= \intprod{H}{\overline{\Omega}} \= Z(\Hv;\Omega)^* \,.
\ee
As indicated in the last 
equality, $\overline{Z}$ is the complex conjugate of the function~$Z$.\footnote{As we discuss in the following sections, $\overline{\Omega}$ and~$\Omega$ 
are no longer complex conjugates in the Euclidean theory and this last equality does not hold in that situation.} This is only true assuming $H$ is real, which is the case in Lorentzian signature. This property will not hold for the Euclidean solutions discussed in Section \ref{sec:new-attractor}.
It is illustrative to write the equation~\eqref{genstabOm} in components. Towards this end, we define 
the normalized fields, of Weyl weight~0,
\be \label{defYG}
\CY^I \= e^{\frac12 \CK(X, \overline{X})} \, \overline{Z}(\Hv;\Omega) \, X^I \,, \qquad 
\CG_I \= e^{\frac12 \CK(X, \overline{X})} \, \overline{Z}(\Hv;\Omega) \, F_I \,, \qquad 
\ee
and similarly~$\overline{\mathcal{Y}}^I$, $\overline{\mathcal{Y}}_I$ by the complex conjugate equations. 
Note that the new fields~$\CY^I$, $\CG_I$ are not holomorphic, and  
are functions of the original period vector as 
well as the sources. 
Then, the components of~\eqref{genstabOm} are given by  
\be \label{genstab}
\i \bigl(  \CY^I - \overline{\CY}^I  \bigr) \= \wt H^I \,, \qquad 
\i \bigl(  \CG_I -  \overline{\CG}_I  \bigr) \= H_I  \,.
\ee

The gauge field is given by
\be \label{FGvals}
\mathcal{A} \= 
e^{2U} \bigl(\overline{Z}(H,\Omega) \Omega + Z(H,\Omega) \overline{\Omega} \bigr) (\dd t+\omega) + \mathcal{A}_d ,
\qquad 
\dd \mathcal{A}_d \= \bast \dd H \,,
\ee
where $ \mathcal{A}_d $ is a gauge field on base space, $ \bast \dd$ is the Hodge dual of the differential form on base space and $H = (\tilde H^I, H_I)$ only depends on the base space coordinate $\xvec$. The field strengths are determined from this gauge field as 
\be 
\label{eq:field-strength-from-gauge-fields}
\mathcal{F} \= (F^I , G_I) \= \dd \mathcal{A} \,.
\ee
The equations of motion and Bianchi 
identities of the gauge fields then impose that~$H_I(\xvec)$ and~$\wt H^I(\xvec)$ 
are \emph{harmonic} functions sourced by the electric and magnetic charges, 
respectively. 

Finally, the metric~\eqref{IWPmetric} is related to the vector multiplet fields 
as follows~\cite{LopesCardoso:2000qm}. The warping of the base space is given by 
\be \label{Uomgeneral1}
 \rme^{-2U}  
\= m_\text{P}^2 \, Z(\Hv;\Omega) \, \overline{Z}(\Hv;\Omega) \,,
\ee
and the connection\footnote{Note that the~$U(1)$ connection of 
the K\"ahler geometry is given 
by $Q_m=\frac12 \, \rme^{2U} \ve_{mnp}\, \p_n \omega_p$.}~$\omega$ is given by 
\be  \label{Uomgeneral2}
\bast\dd\omega \=  \langle \dd\Hv, \Hv \rangle \qquad \Longleftrightarrow \qquad 
\ve_{mnp}\, \p_n \omega_p  
\=  H_I \p_m \wt H^I - \wt H^I \p_m H_I  \,,
\ee
and vanishes for static solutions. 
Consistency of \eqref{Uomgeneral2} with \eqref{genstabOm} further implies that 
the boundary condition for the phase $\alpha(\xvec)$ of $Z(H,\Omega)= e^{-U} e^{\i \alpha}$ at infinity is $\alpha_\infty = \arg(Z(\Gamma,\Omega_\infty))$, where $\Gamma$ is the total charge measured at infinity.\footnote{See the discussion around eqn. (7.16) of \cite{Denef:2000nb}.}

\bigskip

The equations~\eqref{genstabOm}, equivalently~\eqref{genstab},
were first found in the context of the 
spherical supersymmetric black hole.
In this context they were called the attractor equations, and they take a particularly simple form. 
We review this in the following section. 
It is important to note that the generalized stabilization equations are algebraic in nature. In the following presentation 
we denote their solutions by the subscript~$*$ in various functions (or sometimes constant values). 
In particular, the solution for 
the period vector is called~$\Omega_*(H)$. The attractor central charge function is defined to be 
\be \label{Zstardef}
Z_*(H) \; \coloneqq \; Z \bigl( H , \Omega_* (H) \bigr) \= 
\langle H , \Omega_* (H) \rangle 
\= \rme^{\half \CK(X_*, \overline{X}_*)}  \, \bigl(\wt H^I F_{I*} - H_I X_*^I \bigr) \,.
\ee
Finally the solutions for the components are denoted by~$\CY^I_*(H)$,~$\CG_{I*}(H)$. It is easy to see from the definitions and the 
generalized stabilization equations~\eqref{genstabOm},~\eqref{genstab} that 
these functions obey the following scaling properties:
\be \label{OmZscaling}
\begin{split}
& \Omega_*(\lambda H) \=  \Omega_*(H)\,, \qquad 
Z_*(\lambda H) \=  \lambda \, Z_*(H)\,, \\
& \bigl( \,\CY^I_*(\lambda H) \,, \, \CG_{I*}(\lambda H) \bigr) \= \lambda  \, \bigl( \, \CY^I_*(H) \,,\, \CG_{I*}(H) \bigr)\,.
\end{split}
\ee

\bigskip

\subsection{Extremal attractor black hole solutions \label{sec:old-attractor-black-hole}}

The attractor black hole is a particular case of the general family of solutions discussed above. 
It is a regular Lorentzian solution corresponding to the static, spherically symmetric extremal black hole. The harmonic functions are sourced by electric and magnetic charges at the origin, i.e., 
\be
\label{eq:h-and-tilde-h-attractor-solution}
H_I \= h_I + \frac{Q_I}{|\xvec - \xvec_\text{BH}|}\,, \qquad \wt H^I \= \wt h^I + \frac{P^I}{|\xvec-\xvec_\text{BH}|}\,, \qquad  H = h + \frac{\Gamma}{|\xvec-\xvec_\text{BH}|}\,,
\ee
with~$h_I$, $\wt h^I$ being real constants, and $h$ being their associated symplectic vector.
The metric now has the form
\be  \label{sphersymmmetric}
ds^2 \= - \rme^{2U} \dd t^2 +  \rme^{-2U} \dd x^m \dd x^m\,,
\ee
where the function~$U$ depends only on~$\xvec$ and is determined by~\eqref{Uomgeneral1}.
This configuration describes an extremal black hole solution with charges~$(P^I,Q_I)$ 
whose horizon location in base space is at $\xvec = \xvec_{BH}$. 
All the fields of the theory take a very simple, spherically symmetric form near the horizon.
In particular, the period vector $\Omega$ has a constant 
value~$\Omega_*(\Gamma)=\rme^{\half \CK(X_*, \overline{X}_*)} (X^I_*,F_{I*})$ at the horizon,
which can be determined by a consistent application of the generalized stabilization
equations as follows. 

As~$r=|\xvec-\xvec_{BH}| \to 0$ the harmonic function behaves as~$H \sim \frac{\Gamma}{r}$. 
The equations~\eqref{genstab} imply that 
\be
\CY^I \, \sim \, \frac{Y^I_*}{r}  \,, \qquad \CG_I \, \sim \, \frac{G_{I*}}{r} \,, 
\ee
where the constant values~$(Y^I_*, G_{I*})$ obey
\be \label{stab}
\i \bigl( Y^I_*  - \overline{Y}^I_*  \bigr) \= P^I \,, \qquad 
\i \bigl( G_{I*}  - \overline{G}_{I*} \bigr) \= Q_I \,.
\ee
These equations are the original \emph{attractor equations} 
or~\emph{stabilization equations},
that fix the scalars to their \emph{attractor values} at the horizon in terms of the charges. 
If we think of~$Y^I_*$ as~$\nv+1$ independent complex variables, the first set of equations in~\eqref{stab} 
determines the imaginary parts of~$Y^I_*$ to be~$-P^I/2$, and the second set of equations determines
the real parts of~$Y^I_*$ through the (non-linear) dependence of the prepotential on~$Y^I$.

The scaling property~\eqref{OmZscaling} of the attractor central charge function implies that
near the horizon, i.e.~as~$r=|\xvec-\xvec_{BH}| \to 0$,   
\be  \label{Znearhor}
Z_*(H(r)) \; \sim \; \frac{Z_*(\Gamma)}{r}  \,.
\ee
The numerator on the right-hand side, which is a function only of the charges 
\be \label{Zstardef2}
Z_*(\Gamma) \= 
\langle \Gamma , \Omega_* (\Gamma) \rangle 
\= \rme^{\half \CK(X_*, \overline{X}_*)}  \, \bigl(P^I F_{I*} - Q_I X_*^I \bigr) \,.
\ee
is called the \emph{attractor central charge}. The attractor value of the period vector~$\Omega_*$ is related to that 
of the normalized scalars $(Y^I_*,G_{I*})$ as
\be \label{YXrel}
(Y^I_*, G_{I*}) \= (\overline{C} X^I_*, \overline{C} F_{I*}) \,, \qquad 
C \= \rme^{\frac12 \CK(X^I_*, \overline{X}^I_*)}\, Z_*(\Gamma) \,.
\ee

\bigskip

Finally, it follows from~\eqref{Uomgeneral1} that as~$r\to 0$, 
\be \label{USrel}
 \rme^{-2U} \, \sim \, 
    \frac{Z_* (\Gamma) \, \overline{Z}_* (\Gamma) }{r^2}  \,,
\ee
and that the metric~\eqref{sphersymmmetric} takes the form AdS$_2 \times$~S$^2$. 
The Bekenstein-Hawking entropy is now easily calculated by computing the area of the sphere at~$r=0$, 
\be \label{SBH1}
\SBH
\= \pi \, Z_* (\Gamma) \, \overline{Z}_* (\Gamma) \,.
\ee
The right-hand side of this formula makes it manifest that the extremal black hole entropy only depends on the charges.
Another useful expression for the entropy is obtained by combining~\eqref{YXrel}, \eqref{Zstardef}, and~\eqref{SBH1}, 
\be  \label{SBH2}
\SBH 
\= \pi \, \bigl(P^I \, G_{I*} -  Q_I \,  Y_*^I  \bigr) \,. 
\ee
purely in terms of the attractor vaues of the scalars.  
This expression is holomorphic in~$Y^I,G_I$, but note that the rescaling~\eqref{YXrel} 
is not holomorphic in~$X^I,F_I$. A third expression can be obtained from the 
second equation of~\eqref{YXrel} and using the scaling property of the K\"ahler potential~\eqref{defK}, 
\be  \label{SBH3}
\SBH 
\= \pi \i \bigl(Y^I_* \overline{G}_{I*} - \overline{Y}^I_* G_{I*} \bigr)\,,
\ee

\bigskip

The attractor entropy formula for supersymmetric black holes in compactifications of string theory 
agrees with the corresponding string-theoretic microscopic formulas for the index whenever they 
are known~\cite{Sen:2007qy}. 
However, it has not been derived from the gravitational path integral formalism. 
This is precisely what we do in the following sections.
In order to reach that goal, we first discuss more general BPS solutions of~$\CN=2$ supergravity.

\subsection{Spacetime dependence of the general BPS solution \label{sec:spacetime-dep}} 

In the above subsection we saw how the BPS equations discussed in Section~\ref{subsec:BPSeqns}
leads to an elegant description of the extremal BH. In fact, the BPS equations contain a  
remarkable, more general structure~\cite{Behrndt:1997ny, LopesCardoso:2000qm}. 
The equations~\eqref{FGvals}, \eqref{Uomgeneral1}, 
and~\eqref{Uomgeneral2} completely determine the metric and gauge fields in terms of 
the scalar fields (equivalently, the period vector~$\Omega$) and the sources~$H({\mathbf x})$. 
The period vector in turn is determined as the solution to the generalized stabilization equations~\eqref{genstab}. 
These latter equations have exactly the same structure as the stabilization equations~\eqref{stab}
for the fixed scalars at the horizon of the extremal attractor black hole---which are purely algebraic. 
Therefore the solution~$\Omega_*(\Gamma)$ to the extremal attractor stabilization equations gives us 
the spacetime-dependent period vector~$\Omega = e^{\mathcal{K}/2}(X^I(x),F_I(x))$ as~$\Omega=\Omega_*(\Hv(x))$,
and consequently the complete configuration including the metric and gauge fields, as a function of 
the sources~$\Hv=(\wt H^I(x),H_I(x))$ for~\emph{any} BPS solution!  

The general solution can be summarized as follows~\cite{Denef:2000nb, Bates:2003vx}. 
One writes the harmonic function as a sum over a number of sources labelled by~$a$, 
at locations~${\mathbf x}_a$ with charges~$\Gamma_a$
\be \label{eq:Hhh}
H({\mathbf x}) \= \sum_{a} \frac{\Gamma_a}{|\xvec - \xvec_a|} + h \,, 
\qquad h \= - 2 \, \Im(\rme^{-\i \alpha}\Omega)_{r=\infty} \,.
\ee
Here $\alpha|_\infty = \arg(Z(\Gamma;\Omega_\infty))$ is the argument of the central charge at infinity of the 
total charge $\Gamma = \sum_a \Gamma_a$. 
We have chosen the vector of constants~$h$ to be consistent with the 
generalized attractor equations~\eqref{genstabOm} at infinity. 
In Lorentzian signature, each source is interpreted as an extremal black hole 
with monopole charge~$\Gamma_a$ located at the position~$\xvec_a$. 
In Euclidean signature this interpretation will change, as 
we see in the following sections.

The solution for all the fields is determined by the generalized attractor equations 
in terms of a single function of the harmonic sources  
\be
\Sigma(H) \= Z_*(H) \, \overline{Z}_*(H) \,, 
\ee
called the entropy function in~\cite{Bates:2003vx}. 
(Recall that~$Z_*$ is the attractor central charge function defined in~\eqref{Zstardef}.) 
An explicit construction of this entropy function is given for cubic prepotentials in the nice paper~\cite{Shmakova:1996nz}. 
Using~\eqref{Uomgeneral1}, the metric is given by 
\be\label{eq:metricSigma}
\dd s^2 \= - \frac{1}{\Sigma(H)} (\dd t +\omega)^2 + \Sigma(H) \dd x^m\dd x^m \,.
\ee
In order to obtain asymptotic flat space one has~$\Sigma(h)=1$, which is implied by the choice of $h$ in \eqref{eq:Hhh}.
The scalar fields and gauge fields can be determined in terms of the entropy function as follows. 
By taking the intersection product of \eqref{genstabOm} first with $\Omega$ and second with $D_A \Omega \equiv \partial_A \Omega + \frac{1}{2} \partial_A \mathcal{K} \Omega $, 
and using $\langle \Omega  , D_A \Omega\rangle  = \langle \bar{\Omega}, D_A \Omega \rangle = 0$ one finds that $Z_*=\langle H, \Omega \rangle|_{\Omega = \Omega_*(H)}$ and also $(\partial_A e^{-2U} )|_{\Omega = \Omega_*(H)}=(\partial_{\bar{A}} e^{-2U}) |_{\Omega = \Omega_*(H)}  = 0$. Remembering that $\Sigma(H) = e^{-2U}|_{\Omega = \Omega_*(H)}$, we therefore find that the derivative of the $\Sigma$ function is simply given by
\be 
I^{\alpha \beta} 
\frac{\partial \Sigma}{\partial H^\beta} \= \overline{Z}_*(H) \, \Omega_{*}^\alpha(H) 
+ 
Z_*(H) \, \overline{\Omega}_{*}^\alpha(H) 
\,.
\label{eq:entropy_function_derivative}
\ee
This, together with \eqref{genstabOm}, allows one to express the scalar fields as
\be\label{eq:scalar-fields-complexified}
t^A (H) \= \frac{X^A}{X^0} \=
\frac{\i\tH^A + \partial_{H_A}\Sigma}{\i\tH^0 + \partial_{H_0}\Sigma}\,, \qquad 
\overline{t}^A (H) \=  
\frac{\overline{X}^A}{\overline{X}^0} \=
\frac{-\i\tH^A + \partial_{H_A}\Sigma}{-\i\tH^0 + \partial_{H_0}\Sigma} \,.
\ee
Similarly, directly from \eqref{eq:entropy_function_derivative}, the gauge fields can immediately be rewritten as 
\be
\mathcal{A}^{\alpha} \= I^{\alpha \beta} \partial_{\Hv^\alpha} \log (\Sigma) (\dd t+ \omega) + \mathcal{A}_d^\alpha \,, 
\qquad 
\dd \mathcal{A}_d^\alpha \= \bast \dd H^\alpha \,,
\ee
which determine the field strengths from \eqref{eq:field-strength-from-gauge-fields}.

\section{The new attractor mechanism}
\label{sec:new-attractor}

In this section we present and analyze the saddle-points of the index in ungauged \hbox{$\CN=2$} supergravity coupled to vector multiplets.  
These are new supersymmetric Euclidean solutions of the theory which are non-extremal and smooth. 
In Section~\ref{sec:SugraSmoothness} we analyze the constraints imposed by the smoothness of the solution. 
In Section~\ref{sec:NewAttractor} we discuss the new attractor mechanism obeyed by these solutions. 
In Section~\ref{sec:some-examples} we  
illustrate some of the features of these solutions using examples in pure supergravity and a Calabi-Yau compactification.  
We then discuss a particular solution in the general supergravity in 
which the scalars take the attractor values throughout spacetime and the solution reduces to the IWP solution. 
Finally, in Section~\ref{sec:more-general-attractor} we discuss the uniqueness of our solutions to the generalized attractor equations. 

\subsection{Supersymmetric Euclidean black holes in supergravity \label{sec:SugraSmoothness}}

To find the finite temperature supersymmetric solutions of a general supergravity theory, 
we parallel the reasoning presented for Einstein-Maxwell theory in Section~\ref{sec:simplest-saddle-grav-index}.  
We begin with the most general local solution of supergravity that admits Killing spinors~\eqref{IWPmetric}:
\be 
\dd s^2 \= \rme^{2U} (\dd t_E + \omE)^2 + \rme^{-2U} \dd \xvec^2 \,, \qquad \bast \dd\omE \= \i \langle \dd\Hv, \Hv \rangle 
\label{eq:Bates-Denef_metric_Euclidean}
\ee
with $\rme^{2U}= \Sigma (\Hv(\xvec))$  as determined by the generalized stabilization equations for a given symplectic vector~$\Hv$ of harmonic functions.

As reviewed in Section~\ref{sec:old-attractor-black-hole} the harmonic functions~$\Hv(\xvec)$  specifying the 
classic extremal attractor solution all share the same pole location at $\xvec = \xvec_\text{BH}$. 
This results in the emblackening factor $\Sigma(\xvec) = \Sigma[\Hv(\xvec)]$ having a double pole 
at $\xvec = \xvec_\text{BH}$. Just as in Section~\ref{sec:simplest-saddle-grav-index}, this double pole 
implies~$\beta \to \infty$ and is inconsistent  
with a smooth saddle for an index with any finite~$\beta$.\footref{footnote:making-extremal-BHs-periodic}

To obtain the supersymmetric finite temperature black hole solutions, we split the double pole 
in~$\Sigma(\xvec)$ associated to a single black hole into two poles. (This should be reminiscent of \eqref{eq:extremalMetric} being generalized to \eqref{eq:IWPmetric}.) The charge-vector~$\Gamma$
of the black hole correspondingly splits into two charge-vectors,~$\gamma_N$ and~$\gamma_S$,  
associated with the north pole and south pole, respectively, such that the flux measured 
through a closed surface surrounding both the poles yields the total charge of the black hole. 
We define
\be 
\label{chargesplit}
\gamma_N \= \frac{1}2 \left(P^I+\i n^I, Q_I + \i m_I\right)= \frac{\Gamma}{2} + \i \delta \,,\quad 
\gamma_S \= \frac{1}2 \left(P^I-\i n^I, Q_I - \i m_I \right)= \frac{\Gamma}{2} - \i \delta \,,
\ee 
with 
\be 
\gamma_N + \gamma_S \=\Gamma= (P^I , Q_I)\,,\qquad \delta = \frac{1}{2}(n^I ,m_I)\,.
\ee
For now, we place no constraint on the reality of $n^I$ and $m_I$.
The harmonic function of the corresponding solution will therefore now be of the form 
\be \label{HNSsplit}
\Hv(\xvec) \= h + \frac{\gamma_N}{|\xvec-\xvec_N|} + \frac{{\gamma}_S}{|\xvec-{\xvec}_S|}\,, 
\ee
with $\Sigma(h)=1$, which can be compared to the harmonic function \eqref{eq:h-and-tilde-h-attractor-solution} of the extremal solution.

The form of the harmonic function near one of the poles will often appear in the following discussion. Denoting the poles by~$\xvec_a$, $a=N/S$, and the radial distance to the poles by~$\rho_a = |\xvec - \xvec_a|$, 
we have, as~$\rho_a \to 0$, 
\be
H(\xvec) \; \sim \; \frac{\gamma_a}{\rho_a} + 
H^{(0)}_a + \text{O}(\rho_a) \,,  
\ee
where the constant part of this Laurent expansion at the pole is given by 
\be
H^{(0)}_a \= h + \frac{\Gamma -\gamma_a}{|\xvec_N-\xvec_S|} \,.
\ee

By splitting the original center, we have therefore introduced the following new parameters in the solution: the $(2\nv +2)$-component dipole charge vector $\gamma_N - \gamma_S$
and the distance between the north and south poles $|\xvec_N - \xvec_S|$. 
To fix the values of these parameters, we shall once again impose that the geometry should be regular. This means the metric should have no conical singularities and that it should have a finite throat near the horizon in order for the solution to be consistent with a periodic identification of Euclidean time. For the metric written in the form \eqref{eq:Bates-Denef_metric_Euclidean}, these conditions translate to
\begin{itemize} 
\item No conical singularities $\Leftrightarrow$ Singularities of $\omE$ are eliminated through diffeomorphisms
\item \text{No infinite throat/no cusps at the Euclidean horizon} $ \Leftrightarrow $ \\ \text{ }\qquad \qquad $ \Leftrightarrow $
\text{No higher poles near the N/S poles of }$\Sigma(H(\xvec))$.
\end{itemize}

The first condition works essentially as in Section~\ref{sec:simplest-saddle-grav-index}.
We act on the second equation of~\eqref{eq:Bates-Denef_metric_Euclidean} with $\dd \bast $ and integrate the resulting equation 
over a closed volume connecting infinity to a point between the north and south poles, 
and then use Stokes theorem twice. 
The resulting equation when enclosing the north pole is given by
\be 
\label{eq:stokes-multiple-times}
2\pi \, \omEphi^{\hor} \= 
\lim_{\partial S \rightarrow 0}
\int \omEphi \, \dd \phi \;\Bigl{|}_{\rm Dirac ~String} 
\= 4\pi \i \langle \gamma_N, 
\Hv^{(0)}_N 
\rangle
\,,
\ee
implies the existence of a Dirac string singularity running between the north and south poles. 
As in Section~\ref{sec:simplest-saddle-grav-index}, the Dirac singularity can be removed by the coordinate transformation 
$t_E \to t_E' = t_E +  \frac{\i}{\Omega} \phi$,  
where~$\Omega = \i/\omEphi^{\hor}$ is the angular velocity. 
The regularity of the metric (consistency of the periods of angular and time coordinates) leads to the condition \hbox{$\beta \Omega = 2\pi \i$}. 
Upon using \eqref{eq:stokes-multiple-times}, this regularity condition leads to 
\be   \label{eq:regularity_condition_part1}
\i \langle \gamma_N, 
\Hv^{(0)}_N \rangle \=
\frac{\beta}{4\pi}\, .
\ee
Imposing a similar smoothness constraint by enclosing the south pole would lead to an equivalent equation because of the antisymmetry of the intersection product and the relation $\langle \Gamma,h \rangle =0$. The constraint can be written more explicitly as
\be \label{gensmoothcond}
\frac{\i \langle \gamma_N , {\gamma}_S \rangle}{|\xvec_N - {\xvec}_S|} 
\; + \; \i \langle \gamma_N ,h \rangle \= \frac{\beta}{4\pi} \, . 
\ee
The above equation can be viewed as the equation that fixes the distance between the north and south poles in terms of the temperature, charges of the N/S poles, and the asymptotic values of the moduli.
Note that taking the distance between the north and south poles to zero corresponds to taking $\beta \rightarrow \infty$, and equivalently $\Omega \rightarrow 0$, thus recovering the typical extremal solution. 

The second condition implies that the finite temperature black hole does not contain an infinite throat, so that the geometry has the form of a smooth cigar. 
This can be ensured by imposing that the emblackening factor $e^{-2U}$ does not have any second-order poles but rather, following the Kerr-Newman example presented in Section~\ref{sec:simplest-saddle-grav-index}, has first-order poles
\be  
\label{eq:no-higher-order-poles-1}
\lim_{\xvec \rightarrow \xvec_a} |\xvec - \xvec_a|^2 \, 
\rme^{-2U(\xvec)}
\= 0 , 
\qquad 
\lim_{\xvec \rightarrow \xvec_a} |\xvec - \xvec_a| \,  
\rme^{-2U(\xvec)}
\; \neq \; 0 \,.
\ee
for $a=N\text{ or }S$.
The above constraints on the emblackening factor will turn out to only be satisfied if the north/south pole charges $\gamma_N$ and ${\gamma}_S$ are made complex. 
Note that this second regularity condition is automatically satisfied in the previous discussion of the Kerr-Newman black hole, wherein the 
emblackening factor is a product of two harmonic functions $V$ and $\widetilde{V}$ with different poles.

With the above two regularity conditions, the distance between the centers $|\xvec_N - \xvec_S|$ as well as the induced dipole charges, get uniquely fixed in terms of the asymptotic boundary conditions and the temperature. In the following section, we explore the attractor equations for these new supersymmetric finite temperature black hole solutions and show that they can be used to explicitly solve \eqref{eq:no-higher-order-poles-1}. 
Just like in Section~\ref{sec:simplest-saddle-grav-index}, it will turn out that the above regularity conditions play a key role in making the resulting on-shell action consistent with the expected result from a supersymmetric index.

\subsection{The new attractor equations and solutions \label{sec:NewAttractor}}

We now study the structure of the supersymmetric solution corresponding to the 
split~$\Gamma=\gamma_N+\gamma_S$ introduced in the above subsection. 
By analyzing the generalized stabilization equations near 
the poles, we obtain a new form of the attractor mechanism 
that fix the value of the  
scalars at each pole. These attractor equations fix $\gamma_N - \gamma_S$ that is associated
with the dipole fields of the solution. Combined with the smoothness condition discussed
in the previous subsection we find that all the parameters of the new solutions 
are fixed.

In order to find new solutions to the generalized stabilization equations in the Euclidean theory, we allow $\gamma_N$ 
and $\gamma_S$ to have complex components. Note that we still demand that the sum~$\Gamma=\gamma_N+\gamma_S$ is 
equal to the original black hole charges and, in particular, real.
As a consequence of complexifying the charges, the harmonic function $\Hv(\xvec)$, the gauge fields, 
and the scalar fields are also complexified. 
In particular, the vectors $\Omega = e^{\CK/2}(X^I, F_I)$ and $\overline \Omega = e^{\CK/2}(\overline X^I, \overline F^I)$ 
are treated as independent variables. These vectors are determined by the~$\Hv(\xvec)$ as in~\eqref{eq:scalar-fields-complexified}, and 
$\Omega$ and~$\overline{\Omega}$ are complex conjugates if and only if $\Hv(\xvec)$ is real. 
Similarly, the central charge function~$Z(\Hv,\Omega) = \langle \Hv , \Omega \rangle$ and the 
function~$\overline{Z}(\Hv,\Omega) = \langle\Hv , \overline{\Omega} \rangle$ 
entering the generalized stabilization equations are complex conjugates only if $\Hv(\xvec)$ is real.

\medskip

Our task then is to find the solutions of the generalized stabilization equations
\be \label{genstabEuc}
\i (\overline{Z}(\Hv,\Omega)\, \Omega(\Hv) - Z(\Hv,\Omega)\, \overline{\Omega}(\Hv)) \= \Hv \,,
\ee
where the harmonic functions are given by the sources split into the north and south poles~\eqref{HNSsplit}. As in Section~\ref{sec:simplest-saddle-grav-index}, we can perform a diffeomorphism 
such that 
$\xvec_N = (0,0,\alpha)$ and $\xvec_S = (0,0,-\alpha)$. 
In the Euclidean theory, $\Omega(H)$ and~$\overline{\Omega}(H)$ are, a priori, two 
independent complex functions.  
Finding all smooth solutions to the stabilization equations with this reality condition is a non-trivial problem.

We proceed for the moment by 
assuming that the solutions are analytic continuation of solutions in the Lorentzian theory 
(even though the Lorentzian solutions may have naked singularities or unacceptable causal properties). 
In practice, the condition of analytic continuation from the Lorentzian theory means that any solution, upon the inverse Wick rotation~$\alpha=-\i a$, should go to a solution with good reality properties in the Lorentzian theory. In particular, the harmonic functions appearing in the BPS solutions should be real.
Since the sources are placed at $\pm \alpha$ on the $z$-axis, this condition 
says that the exchange~$\xvec_N \leftrightarrow \xvec_S$ implements 
the conjugation $\Omega_* \leftrightarrow \overline{\Omega}_*$ in the solution. 
This implies that all the components of $\gamma_N$ and~$\gamma_S$ are complex conjugates of each other and,
in particular, the dipole charges~$\i m_I$ and~$\i n^I$ are purely imaginary.
These assumptions will lead us to a unique solution for the saddle of the index. We do not have a general proof that a solution with these properties is the only solution possible. For a proof that applies only to a theory with $n_V=1$ and a specific prepotential, see Section~\ref{sec:more-general-attractor}.

\bigskip

Consider the behavior of the solution close to one of the poles, say the north pole. 
Choosing a local radial coordinate~$\rho_N=|\xvec-\xvec_N|$, the metric is well-approximated 
as~$\rho_N \to 0$ by the spherically symmetric form~\eqref{sphersymmmetric} 
of the classic attractor mechanism with~$r$ replaced by~$\rho_N$. 
In particular, following the steps leading to~\eqref{USrel}, we obtain 
\be  \label{Znearpole}
Z_*(H(r)) \= \frac{Z_*(\gamma_N)}{\rho_N}  + 
\langle H^{(0)}_N, \Omega_*(\gamma_N) \rangle  + \text{O}(\rho_N) \,, 
\qquad \rho_N \to 0 \,.
\ee 
Here the coefficient of the singular term~$Z_*(\gamma_N) = \langle \gamma_N,\Omega_*(\gamma_N) \rangle$ as well as the constant term 
can be calculated easily using the formula~$Z_*(H) = \langle H,\Omega_*(H) \rangle$ and the 
scaling properties~\eqref{OmZscaling}. 
Consequently, the emblackening factor is, as $\rho_N \to 0$,
\be
\label{eq:emblackening-around-pole}
\rme^{-2U} \= \frac{1}{\rho_N^2}
\bigl(Z_*(\gamma_N) + \rho_N\langle H^{(0)}_N, \Omega_*(\gamma_N) \rangle \bigr) \, 
\bigl(\overline{Z}_*(\gamma_N) 
+ \rho_N \langle H^{(0)}_N, \overline{\Omega}_*(\gamma_N) \rangle \bigr) + \text{O}(\rho_N)
.
\ee
However, as we discussed above, the cigar has a cap, which means that the metric 
should not have a double pole as in the extremal case.
This seems to pose a puzzle: how is the existence of a 
source at~$\mathbf x=\mathbf  x_N$  consistent with the absence of a double pole? 
The resolution is to demand that\footnote{One could also have analogous relations 
with the roles of $Z_*(\gamma_N)$ and~$\overline{Z}_*(\gamma_N)$ reversed.}
\footnote{In the Lorentzian theory it has been argued that configurations with $Z=0$ in the interior of moduli space 
is related to some degeneration like the 
the condensation of massless scalars~\cite{Denef:2000nb}. In contrast, in the Euclidean theory discussed here, we have a regular, weakly-curved region near the horizon.}
\be  \label{Zcenter}
Z_*(\gamma_N) \=  \langle \gamma_N , \Omega_*(\gamma_N) \rangle \= 0 \,, 
\qquad
\overline{Z}_*(\gamma_N) \= \langle \gamma_N , \overline{\Omega}_*(\gamma_N) \rangle \; \neq \; 0 \,. 
\ee
The resulting single pole at $\xvec_N$ that can be read off from~\eqref{eq:emblackening-around-pole}
leads to a spacetime that smoothly caps off and has no infinite throat.  
Such a configuration is only possible if $\overline Z_*(\gamma_N)$ is no longer the complex 
conjugate of~$Z_*(\gamma_N)$, which is only possible if the components of~$\gamma_N$ and~$\gamma_S$ 
are complex.\footnote{The configuration $Z_*(\gamma_N) = \overline Z_*(\gamma_N) = 0$ leads to a trivial 
solution in which the emblackening factor has no pole at all.} 
Using the assumption that the index solution is a continuation from a real Lorentzian solution, at the south pole we have the conjugate equations\footnote{Without this assumption, we could consider $\langle \gamma_a, \Omega_*(\gamma_a) \rangle =0$ at both poles. We will rule this option out in Section~\ref{sec:more-general-attractor}.}
\be  \label{Zcenter2}
Z_*(\gamma_S) \=  \langle \gamma_S , \Omega_*(\gamma_S) \rangle \neq 0 \,, 
\qquad
\overline{Z}_*(\gamma_S) \= \langle \gamma_S , \overline{\Omega}_*(\gamma_S) \rangle \; \= \; 0 \,. 
\ee

Now we look at the behavior of the generalized stabilization equations~\eqref{genstabEuc} close to the poles. 
Upon matching the leading singular behavior of both sides of~\eqref{genstabEuc} as~$\xvec \to \xvec_a $ for $a=N$, $S$, and using~\eqref{Zcenter} and \eqref{Zcenter2} we obtain
\be \label{Poleattr}
\text{North pole: } \quad  \i\overline{Z}_*(\gamma_N) \,\Omega_*(\gamma_N) \= \gamma_N  \,,\qquad \quad 
\text{South pole: } \quad-\i Z_*(\gamma_S) \, \overline{\Omega}_* (\gamma_S) \=  \gamma_S  \,. 
\ee
These are the new attractor equations at the poles. 
Let us look at the north pole equation in terms of the scale-invariant components 
$\overline{Z}(\gamma_N)\Omega(\gamma_N)= (Y^I_N, G_{IN})$. We have 
\be \label{Poleattrcomp}
\i Y^I_N  \= P^I - \i n^I \,,  \qquad  
\i G_{IN}  \= Q_I - \i  m_I \,.
\ee
The real parts of these equations are
\be \label{NPattrcomp}
\i \bigl( Y^I_N - (Y^I_N)^* \bigr) \= P^I  \,,  \qquad  
\i \bigl( G_{IN} -  (G_{IN})^* \bigr) \= Q_I  \,.
\ee 
which are precisely the the classic attractor equations for the extremal black hole with charges~$\Gamma=(P^I, Q_I)$!
The solutions are therefore given by the extremal attractor values~$Y_*$ in~\eqref{stab}.
The same discussion clearly applies to the south pole, where denoting $Z_*(\gamma_S)\overline{\Omega}_*(\gamma_S)=(\overline{Y}_S^I,\overline{G}_{IS})$, we have
\be \label{SPattrcomp}
-\i \bigl( \overline{Y}^I_S - (\overline{Y}^I_S)^* \bigr) \= P^I  \,,  \qquad  
-\i \bigl( \overline{G}_{IS} -  (\overline{G}_{IS})^* \bigr) \= Q_I  \,,
\ee 
which is the complex conjugate of the north pole equations. The solutions are
\be
\begin{split}
Y^I_N  \= Y^I_* \,,& \qquad 
\overline{Y}^I_N  \= 0 \,, \\ 
Y^I_S \= 0\,,& \qquad \overline{Y}^I_S \= 
\bigl( Y^I_* \bigr)^*  \,.
\end{split}
\ee

Now we turn to the imaginary part of the new attractor equations~\eqref{NPattrcomp} at the poles. We obtain
\be \label{NPattrcomp2}
Y^I_N + (Y^I_N)^*  \= n^I  \,,  \qquad  
 G_{IN} +  (G_{IN})^* \= m_I  \,,
\ee 
In other words, the dipole charges are the real parts of the scalars. Since the complete complex scalars are determined by the attractor equations, we see that the dipole charges are indeed determined in terms of the monopole charges, as we mentioned at the beginning of the subsection.

\bigskip

We can summarize this above discussion as follows. 
The old attractor equations at the horizon split into a complex part at the north pole and the 
complex conjugate part at the south pole, as in~\eqref{Poleattr}. Each solution can be 
identified with the old attractor solution, so that we have 
\be
\begin{split}
\label{eq:new_attractor_charges}
\gamma_N &\= \i\overline{Z}_*(\gamma_N) \,\Omega_*(\gamma_N) 
 \= \i\overline{Z}_*(\Gamma)\,\Omega_*(\Gamma)  \,, \\
 \gamma_S &\= -\i Z_*(\gamma_S) \, \overline{\Omega}_* (\gamma_S) 
 \= -\i Z_*(\Gamma) \, \overline{\Omega}_* (\Gamma) \,.
\end{split}
\ee 
The scalars $Y$ and $\overline Y$ are each attracted to their attractor value but at different locations in the spacetime. $X$ gets attracted to its attractor value that depends only on the charges of the black hole at the north pole of our Euclidean solution (with $\overline X$ dependent on the value of the scalar moduli at infinity at this location), while $\overline X$ gets attracted to its attractor value at the south pole of our Euclidean solution (this time, with $X$ dependent on the value of the scalar moduli at infinity at this location). 
Moreover, we have thus used the equations of supersymmetry and smoothness of the spacetime to 
determine all the parameters of our saddle-point solutions.

\subsection{Some  examples}
\label{sec:some-examples}
In this subsection we illustrate the finite-temperature supersymmetric solutions discussed in the previous subsection through simple examples. 
We first discuss pure supergravity. Although it is essentially trivial as far as the dynamics of scalars are concerned, this example is useful to complete the link between the discussions of Sections~\ref{sec:simplest-saddle-grav-index}  and~\ref{sec:sugrarev}. We then discuss a more involved model with a cubic prepotential, which captures Type IIA compactification on a Calabi-Yau manifold with D0-D4 charges. 

\subsection*{Pure supergravity}

This is the theory discussed in Section~\ref{sec:sugrarev} with~$\nv=0$ and the prepotential
\be
F \= \frac{\i}{2} (X^0)^2 \qquad \Longrightarrow \qquad \rme^{-\CK}\=2 X^0 \overline{X^0} \,.
\ee
In order to obtain Poincar\'e supergravity, one sets the right-hand side to a constant as in~\eqref{gaugefixingcondition}. The resulting theory is precisely the 
Einstein-Maxwell theory discussed in
Section~\ref{sec:simplest-saddle-grav-index}. 
It follows directly from the definition of the normalized scalars defined in~\eqref{defYG} that they obey  
the equations 
\be \label{AttrPuresugra}
\i( \CY^0 - \overline \CY^0) \= \wt H^0  \,, \qquad \i(\CG_0 - \overline \CG_0) \= \CY^0 + \overline \CY^0 \=  H_0 \,.
\ee

We first discuss the Lorentzian theory. 
Recall that~$\wt H^0$, $H_0$ here are real harmonic functions, and that the idea of the IWP solutions is to think of these harmonic functions as real or imaginary parts of complex harmonic functions. 
In the supergravity context, the equations~\eqref{AttrPuresugra} make it clear that~$\wt H^0$ and~$H_0$ 
are precisely the real and imaginary parts, respectively, 
of the complex harmonic function~$\CY^0$.
The solution of interest is given by taking~$\CY^0$ to have a source of strength~$Q_0$ 
at~$\xvec_N=(0,0,-\i a)$ with~$a \in \IR$,  and, correspondingly,~$\overline \CY^0$ to have a 
complex conjugate source\footnote{As we see below, $Q$ is exactly the electric 
charge of the solution, in accord with the discussion of the 
electric Kerr-Newman black hole.
More generally, the source of the complex harmonic function is 
$Q_0+ \i P^0$ in terms of the electric and magnetic charges 
as measured from infinity.} at~$\xvec_S=(0,0,\i a)$, i.e.,
\be
\CY^0 \= 1 + \frac{Q_0}{R} \,, \qquad \overline{\CY}^0 \= 1 + \frac{Q_0}{\overline R} \,, 
\ee
with~$R=\sqrt{x_1^2+x_2^2+(x_3+\i a)^2}$ and~$\overline{R}=\sqrt{x_1^2+x_2^2+(x_3-\i a)^2}$. 
The square roots are defined with a  choice of branch cut such that~$R$ and~$\overline{R}$ are complex conjugates. 
This illustrates the idea that the real harmonic functions entering these solutions are taken to be real and imaginary parts of a complex harmonic function. 
Indeed, the harmonic functions~$H_0$ and~$\wt H^0$ can be read off to be 
\be 
H_0 \= 1 + \frac{Q_0}{2 R} + \frac{Q_0}{2 \overline{R}}   \,, 
\qquad \wt H^0 \= \frac{Q_0}{2\i R} - \frac{Q_0}{2\i \overline{R}} \,,
\ee
which illustrates the complex conjugate nature of the solution upon exchange of the sources.
Note that the angular momentum in the Lorentzian theory is~$J=aM$ 
and so the imaginary position of the source is consistent with the reality of angular momentum. 

All this is exactly in accord with the general discussion
of~$\half$-BPS solutions in $\CN=2$ supergravity. 
The Euclidean solution follows in a straighforward manner by the Wick rotation~$a = - \i \alpha$. 
We see immediately that, upon identifying~$\CY^0=V$, $\overline{\CY^0}=\wt V$, we recover precisely the 
IWP solution~\eqref{eq:IWPmetric}--\eqref{eq:choice-of-V-tilde-V}. Finally let us emphasize that this is also a solution of $\mathcal{N}=2$ supergravity coupled to hypermultiplets only. 

\bigskip

\subsection*{Calabi-Yau compactification with D0-D4 charges} 

We now consider an example in the language of string theory\footnote{For a pedagogical introduction we highly recommend \cite{VandenBleeken:2008tsa,Cheng:2008gx}}. We consider Type~IIA supergravity in 10 dimensions compactified on Calabi-Yau 3-fold~$X$. 
The massless field content of the resulting four-dimensional low energy effective theory is then completely determined by the cohomology of the Calabi-Yau manifold. The effective theory admits a truncation, after setting hyperscalars to constant values, which 
is described by the 
four-dimensional $\mathcal{N}=2$ supergravity 
action~\eqref{eq:Lorentzian_action}. The $(2n_V+2)$-component symplectic vectors $A=(\tilde{A}^I , A_I)$ are elements of the even cohomology $H^{2*}(X) = H^0(X) \oplus H^{(1,1)}(X) \oplus H^{(2,2)}(X) \oplus H^{6}(X)$ of respective dimensionalities $(1,h_{1,1}=n_V,h_{1,1}=n_V,1)$. In particular, each vector can be written in a symplectic basis $(\alpha_I , \beta^I)$ as
\be 
A \= (\tilde{A}^I , A_I) \= \tilde{A}^I \alpha_I + A_I \beta^I \,, 
\ee
where $(\alpha_I , \beta^I)= (\alpha_0, \alpha_A ,\beta^A, \beta^0)$ are a harmonic basis of the cohomologies, together forming a symplectic basis with respect to the intersection product 
\be 
\langle \alpha_I , \beta^J \rangle \= 
\int_X \alpha_I \wedge \beta^{J *} \= \delta^J_I \,. 
\ee
Here we defined the operation $E^* \equiv (-1)^n E $ for $E \in H^{(n,n)}(X)$ required to make the above product antisymmetric. A black hole of charge  
\be 
\Gamma \= P^0 \alpha_0 + P^A \alpha_A + Q_A \beta^A + Q_0 \beta^0 , 
\ee
can be interpreted as arising from a D-brane configuration with $P^0$ D6 branes wrapped around the full Calabi-Yau, $P^A$ D4 branes wrapped on four cycles of $X$, $Q_A$ D2 branes wrapped on two cycles of $X$ and $Q_0$ D0 branes located at a point on the Calabi-Yau manifold, 
all positioned at a single point of the four-dimensional flat space.

In the large volume limit of the Calabi-Yau, the prepotential is given by
\be
F \= \frac{1}{6}D_{ABC} \frac{ X^A X^B X^C}{X^0} \,,
\ee
where the completely symmetric tensor~$D_{ABC}$, $A,B,C = 1,\dots , \nv$ 
is given in terms of the intersection numbers of cycles in the Calabi-Yau. 
For simplicity, we focus on the diagonal case where the scalar fields $X^A$ are set to be equal
\be 
\label{eq:diagonal-prepotential}
F \; \equiv \; \frac{d_1}{6} \frac{(X^1)^3}{X^0} , 
\ee
and the resulting action is described by a single vector multiplet $n_V = 1$.

In this case, the generalized attractor equations can be explicitly solved analytically, following \cite{Shmakova:1996nz}. The resulting entropy function takes the form
\be
\label{eq:Sigma(H)-diagonal-prepotential}
\Sigma(H) \= 
\sqrt{
\frac{1}{3} p^2 q^2 - \frac{8}{9 d_1} q^3 v +
\frac{2 d_1}{3} p^3 u - 2 p q u v - u^2 v^2 
} ,
\ee
where, to match the example in \cite{Bates:2003vx}, we adopted a new notation for components of $H(\xvec)$ in $(D6,D4,D2,D0)$-basis
\be
\Hv(\xvec) \= (\wt H^0 , \wt H^1, H_1, H_0) 
\; \equiv\;  (v, p, q, u) \,,
\ee
The scalar field given by~\eqref{eq:scalar-fields-complexified} is 
\be 
\frac{X^1}{X^0} \= t^1 \; \equiv \; \tau \= \frac{pq + 3uv}{d_1 p^2 - 2q v} + \i \frac{3 \Sigma(H)}{d_1 p^2 - 2q v} \,,
\ee
in terms of which the normalized period vector takes the form 
\be   
\Omega \= 
\frac{1}{\sqrt{\frac{4d_1}{3}(\Im \tau)^3}} 
\left( 
-1, -\tau , -\frac{d_1}{2} \tau^2 , \frac{d_1}{6} \tau^3
\right)
.
\ee

As a concrete example, we consider a single-centered D4-D0 brane solution with charges $\Gamma= (P^0, P^1 , Q_1, Q_0) \equiv (0,\hat{p},0,\hat{u})$ and the asymptotic value of the scalar field at infinity $\tau|_{r=\infty} \equiv \i a_\infty$, $a_\infty \in \IR$. 
Let us first recall what the zero temperature solution with such boundary conditions looks like. The central charge at infinity is 
\be   
Z(\Gamma;\Omega_\infty) \= 
\frac{d_1}{2} \hat{p} \, a_\infty^2 + \hat{u}
\,,
\ee
and therefore
\be    
h \; \equiv \;  
-2 \, \Im(e^{-\i\alpha} \Omega)_{r=\infty}
\=
\left( 0,
\sqrt{\frac{3}{d_1 a_\infty}}, 
0, 
\frac{1}{2} \sqrt{\frac{d_1 a_\infty^3}{3}}
\right)
\,,
\ee
where $\alpha = \arg Z(\Gamma;\Omega_\infty) = 0$. 
The harmonic function $H_{\text{extremal}}(x)$ of the zero temperature solution is then
\be   
\Hv_\text{extremal}(x) \= 
\left(0, h_p + \frac{\hat{p}}{\rho} 
,0, h_u + \frac{\hat{u}}{\rho} \right) , 
\qquad h_p \= \sqrt{\frac{3}{d_1 a_\infty}} , \qquad  h_u \= \frac{1}{2} \sqrt{\frac{d_1 a_\infty^3}{3}} \, ,
\ee
where one can explicitly verify $\Sigma(h)=1$.
The entropy of the black hole is found to be 
\be 
\SBH \= 2\pi \sqrt{\frac{d_1}{6}} \sqrt{\hat{p}^3 \hat{u}} .
\ee

We now turn on the temperature for this solution and impose the regularity conditions. The single center now splits into north and south poles, which carry additional dipole charges fixed by the new attractor \eqref{eq:new_attractor_charges} to 
\begin{align}
\i \delta \= \frac{1}{2}(\gamma_N - \gamma_S)
&\=\frac{\i}{2}(\bar{Z}_*(\Gamma) \Omega_* (\Gamma) + Z_*(\Gamma) \overline{\Omega}_* (\Gamma) )
\\
&\=
\left( 
-\frac{\i}{2} \sqrt{\frac{d_1}{6}} \sqrt{\frac{\hat{p}^3}{\hat{u}}} \,, \,
0 \,, \, 
 \frac{\i}{2} \sqrt{\frac{3 d_1}{2}} \sqrt{\hat{p} \hat{u}}
\,,\, 0
\right)
\,.
\label{eq:dipole-charges-example}
\end{align}

\begin{figure}[t!]
    \centering
\includegraphics[width=0.85\textwidth]{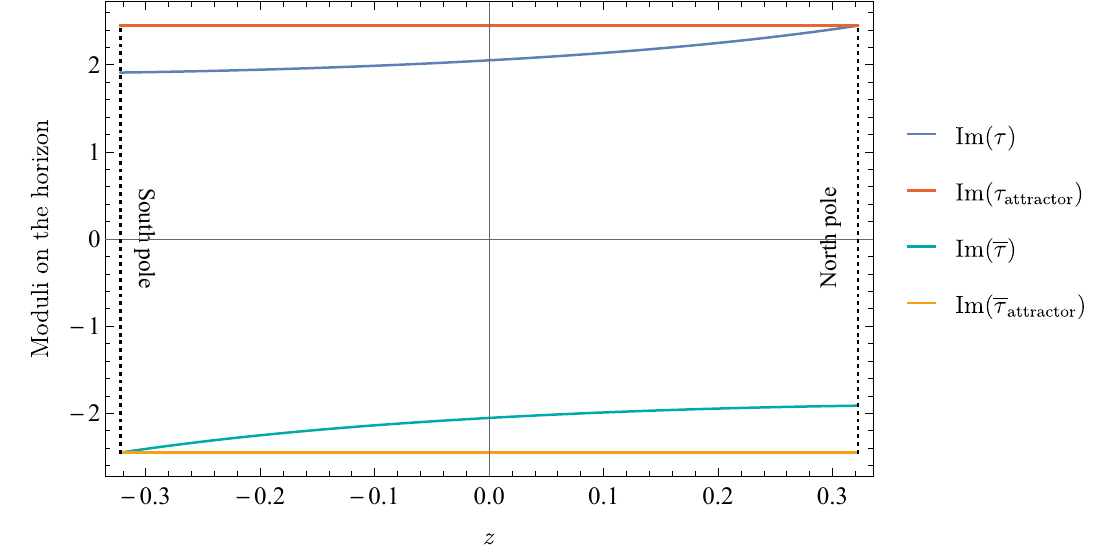}
    \caption{The gauge-invariant scalar ratio $\tau = X^1/X^0$ and $\bar \tau = \bar X^1/\bar X^0$ along the horizon of the black hole (which we place on the $z$-axis), between the south and north poles. In this example, we fix the asymptotic moduli to be purely imaginary in which case the $\tau$ and $\bar \tau$ are purely imaginary. As expected, $\tau$, whose imaginary part is shown in blue, gets attracted to its attractor value, shown by the red vertical line, at the north pole. Similarly, $\bar \tau$, whose imaginary part is shown in cyan, gets attracted to its attractor value, shown by the yellow vertical line, at the south pole. Away from these two points $\tau$ and $\bar \tau$ no longer flow to their attractor value and instead have a non-trivial profile along the horizon. }
    \label{fig:scalar-profile-D0-D4}
\end{figure}

Thus we see that the finite temperature supersymmetric saddle develops imaginary D6-D2 dipole charges.
The harmonic function $H(x)$ of the new finite temperature solution now takes the form
\be 
\label{eq:Hamronic-function-example}
\Hv \= 
\left(
\frac{\i n^0}{2\rho_N} - \frac{\i n^0}{2\rho_S},
h_p + \frac{\hat{p}}{2\rho_N} + \frac{\hat{p}}{2\rho_S} ,
\frac{\i m_1}{2\rho_N} - \frac{\i m_1}{2\rho_S},
h_u + \frac{\hat{u}}{2\rho_N} + \frac{\hat{u}}{2\rho_S} 
\right)
\,,
\ee
where we wrote the dipole charge in components $\delta \equiv \frac{1}{2}(n^I , m_I)$.
The distance between the north and south poles is now fixed by regularity to be
\be  
\i \,  \langle \gamma_N , H^{(0)}_N
\rangle
\= \frac{\beta}{4\pi} ,
\ee
which explicitly takes the form ($h \equiv (0 , h_p ,0 , h_u)$) 
\be   
\frac{ \i \langle \gamma_N , \gamma_S \rangle}{|\xvec_N - \xvec_S|} + \i \langle \gamma_N ,h \rangle \= 
\frac{\beta}{4\pi} \qquad
\Rightarrow \qquad
\frac{ \hat{p}  m_1-  \hat{u}  n^0}{2|\xvec_N - \xvec_S|} 
 - \langle \delta, h \rangle
\= \frac{\beta}{4\pi} \,.
\label{eq:D0D4-regularity}
\ee
From the harmonic function \eqref{eq:Hamronic-function-example} together with the dipole charges \eqref{eq:dipole-charges-example}, the profile for the moduli $\tau = X^1/X^0$ and $\bar \tau = \bar X^1/\bar X^0$ can be computed on the entire manifold. To emphasize that, in contrast to the standard attractor mechanism, the moduli are no longer constant on the horizon, we show their profile along the $z$-axis in figure \ref{fig:scalar-profile-D0-D4}.

To gain more intuition for the new solution, let us analyze its angular momentum. This can be read off from the falloff of $\omega_E$ at infinity through
\be    
\omega_E \= 2 \i  \epsilon_{ijk} \frac{J^i x^j dx^k}{r^3} + O(r^{-2}) .
\ee
From equations for $\omega_E$, asymptotically we have 
\be  
\bast \dd \omega_E \=\i \, \langle \dd\Hv ,\Hv \rangle 
\; \simeq \; 
\dd \left[ \frac{\langle h, \delta \rangle (\xvec_N - \xvec_S) \cdot \xvec}{r^3} \right] + \dots ,
\ee
from which we find 
\be  
J \= 
\frac{\i}{2} \langle h,\delta \rangle 
(\xvec_N - \xvec_S )
\=
\frac{\i}{4}(h_p m_1- h_u n^0) (\xvec_N - \xvec_S)
\,.
\ee
Using now the regularity condition \eqref{eq:D0D4-regularity} this can be rewritten as 
\be 
J = \frac{\i}{2} \langle \delta , \Gamma \rangle \frac{\xvec_N -\xvec_S}{|\xvec_N - \xvec_S|} + 
\i \frac{\beta}{8\pi} (\xvec_N - \xvec_S) .
\ee
Apart from the topological part of the angular momentum captured by the first term, the angular momentum also contains a term that is proportional to the distance between the poles.
This therefore differentiates this two-centered solution from the bound state solutions of~\cite{Denef:2000nb} where the angular momentum is purely topological. 

The explicit form of the angular momentum allows us to complete the analogy with the supersymmetric Kerr-Newman solution. Recalling that the area of the finite temperature horizon is given by 
\be 
A_\text{H} \= 2\pi \,   \abs{\xvec_N - {\xvec}_S} \left(\omega_{E \,\phi}|_{\text{hor}} \right) 
\= \beta \abs{\xvec_N - {\xvec}_S} \,, 
\ee
and that the regularity of the metric also fixes $\beta \Omega = 2\pi \i$, 
the condition \eqref{eq:D0D4-regularity} can be rewritten as  
\be    
-\frac{\pi \i}{2} \, (\i \hat{p}m_1- \i \hat{u}n^0) \=
\beta \Omega J + \frac{A_\text{H} }{4} \, .
\ee
Plugging in the explicit values of the dipole charges the condition takes the form
\be 
\beta \Omega J + \frac{A_\text{H} }{4} \= 2\pi \sqrt{\frac{d_1}{6}} \sqrt{\hat{p}^3 \hat{u}} \= \SBH
\,, 
\ee
which can be recognized as the quantum statistical relation for the index. 
Furthermore, one can now explicitly evaluate the on-shell action for this finite temperature supersymmetric solution to find 
\be 
- I_{E}^{\text{on-shell}} \= -\beta |Z(\Gamma,\Omega_\infty)| + \SBH \,.
\ee
Thus, imposing the regularity of the finite temperature geometry fixes the value of its on-shell action to be consistent with the supersymmetric index. 
As we will see in the following section, this conclusion extends beyond the example considered here. In particular, we show that this conclusion persists for the cases with any number of vector multiplets $n_V$ and black holes with general charges.

\subsection*{Reduction to the Israel-Wilson-Perj\'es Solution} 

In the first example of this subsection we saw how the supersymmetric rotating black hole solution of $\mathcal{N}=2$ supergravity found in this paper reduces to the IWP solution in the absence of vector multiplets. 
Now we show a similar reduction takes place in the presence of vector multiplets for specific choices of asymptotic moduli. This fact was assumed implicitly in \cite{Iliesiu:2021are} and more recently \cite{Iliesiu:2022kny} and \cite{H:2023qko}. 

As a warm up we shall consider the extremal (Lorentzian) case first. The general extremal multi-center solution is well-known. It is a multi-center generalization of the  solution found  in  Section~\ref{sec:NewAttractor} (simply promote the harmonic function to a sum over multiple centers), but allows for the presence of double poles in $\Sigma(H)$ at each center since we begin by considering the extremal case. The purpose here is to find under which conditions the general multi-center solution reduces to the Majumdar-Papapetrou solution described in Section~\ref{sec:simplest-saddle-grav-index}. We can take the source functions $H$ to be 
\begin{equation}
    H \= h + \sum_a \frac{\Gamma_a}{|\xvec-\xvec_a|}\,,
\end{equation}
where the label $a$ runs over the location of the black hole horizons. Now we make the first restriction; take all the charge vectors to be parallel, i.e., 
\begin{equation}
\Gamma_a \= \Gamma \, {\sf e}_a \,, \qquad \sum_a {\sf e}_a \= 1\,,
\end{equation}
where $\Gamma$ is the total charge vector 
and ${\sf e}_a$ are numbers (not vectors) associated to each center. 
The second restriction concerns the boundary conditions for scalar fields at infinity.
Consider the case that $h$ is proportional to the same charge vector $h = \Gamma {\sf h}$, where ${\sf h}$ is a real determined by the asymptotic boundary conditions of the metric. With these restrictions the harmonic vector becomes
\begin{equation}
    H \= \Gamma \underbrace{\left( {\sf h} + \sum_a \frac{{\sf e}_a}{|\xvec-\xvec_a|}\right)}_{\equiv \lambda(\xvec)}\,.
\end{equation}
We can now exploit the scaling properties listed in~\eqref{OmZscaling}. Since the attractor equations are algebraic, the fact that the function $\lambda(\xvec)$ depends on spatial position is irrelevant, and we can use the relations \eqref{OmZscaling} locally at each spacetime point. In particular 
\begin{equation}
    \Omega_*(\Gamma \lambda(\xvec)) \= \Omega_* (\Gamma)\,, \quad \bigl( \,\CY^I_*(\Gamma \lambda(\xvec)) \,, \, \CG_{I*}(\Gamma \lambda(\xvec)) \bigr) \= \lambda(\xvec)  \, \bigl( \, \CY^I_*(\Gamma) \,,\, \CG_{I*}(\Gamma) \bigr)\,.
\end{equation}
These relations mean that the scalars are given by $X^I = \lambda(\xvec) X^I_*$, where $X^I_*$ correspond to the attractor values near each center. Since all $\Gamma_a$ are proportional to $\Gamma$, the attractor value at each center takes the same value, up to a possible overall rescaling. 
Therefore we conclude that even though the scalars $X^I$ can vary in space, the whole space dependence is an overall factor that cancels when computing the physical variables 
$t^A(\xvec) = X^A/X^0 = t^A_*$, 
which take the attractor value everywhere. 
(We see why having parallel charges is crucial; otherwise, the attractor values in each center would differ and having a solution with homogeneous $t^A$ would be impossible.) 
Finally, we can check that the metric takes the form~\eqref{eq:extremalMetric}. First notice that all charges being parallel implies that $\omega=0$. Second, using the remaining relation in \eqref{OmZscaling}, and setting ${\sf h } = 1/|Z_*(\Gamma)|$ we can evaluate the emblackening factor
\begin{eqnarray}
Z_*(\Gamma \lambda(\xvec)) &\=&  \lambda(\xvec) \, Z_*(\Gamma)\,\nonumber\\
\Rightarrow~~ \Sigma(\xvec) &\=& Z_*(\Gamma \lambda(\xvec)) \overline{Z}_*(\Gamma \lambda(\xvec)) \=  \left(  1 + \sum_a {\sf e}_a\,\frac{|Z_*(\Gamma)|}{|\xvec-\xvec_a|} \right)^2 \,,
\end{eqnarray}
which, thanks to the choice of ${\sf h}$, satisfies $\Sigma(\xvec\to\infty)=1$. Inserting this expression for $\Sigma$ in~\eqref{eq:metricSigma} we recover precisely \eqref{eq:extremalMetric} with $V^2 = \Sigma$. 
Moreover we can interpret the coefficients ${\sf e}_a$ as a choice specifying how the total charge $|Z_*(\Gamma)|$ is distributed between each center. Therefore the Majumdar-Papapetrou solution, derived from a theory without vector multiplets, applies to any theory of supergravity as long as the moduli at infinity are set to its attractor value and as long as all charges are parallel.

\medskip

We shall next consider the case of interest in this section; the finite temperature, rotating and supersymmetric black hole with total charge $\Gamma$. 
We will show that when the moduli at infinity are set to the attractor value associated to $\Gamma$, $\Omega_\infty = \Omega_*(\Gamma)$, the solution reduces to the IWP solution of Section \ref{sec:simplest-saddle-grav-index}, equivalent to Kerr-Newman. Introduce the following decomposition 
\begin{equation}
    H \= H_N + H_S \,, \qquad H_a \= h_a + \frac{\gamma_a}{|\xvec-\xvec_a|}\,,
\end{equation}
such that both $\Gamma= \gamma_N + \gamma_S$ and $h=h_N + h_S$. We consider known the new attractor solution, meaning the values of $\gamma_{N/S}$ and $\Omega_*/\overline{\Omega}_*$. We make the following choice; take a solution where $h_a = \gamma_a {\sf h}_a$, where ${\sf h}_a$ is a number (similar to the case considered above).
Next, choose an ansatz for $\Omega$ and $\overline{\Omega}$ that is constant throughout space and satisfies the new attractor equations $ \langle \gamma_N, \Omega_* \rangle = 0$ and $\langle \gamma_S , \overline{\Omega}_* \rangle = 0$. This implies that at every point in spacetime 
\begin{eqnarray}
    Z(H,\Omega_*) &\=&\langle H_N + H_S, \Omega_* \rangle \= Z(H_S,\Omega_*) \,,\nonumber\\
    \overline{Z}(H,\Omega_*)&\=&\langle H_N + H_S , \overline{\Omega}_* \rangle \= \overline{Z}(H_N,\Omega_*) \,.\label{rere}
\end{eqnarray}
We can show that this ansatz solves the generalized stabilization equations, which can be decomposed as
\begin{equation}\label{sasa}
    \i (\overline{Z}(H_N+H_S,\Omega_*) \Omega_* - Z(H_N+H_S,\Omega_*) \overline{\Omega}_*) \= H_N + H_S \,.
\end{equation}
We use that $H_a = \gamma_a \lambda_a (\xvec)$, where $\lambda_a (\xvec)={\sf h}_a + 1/|\xvec-\xvec_a|$ are scalar functions, to show the stabilization equations are satisfied, as follows. Since we chose a constant $\Omega_*$ and $\overline{\Omega}_*$ such that $\i \langle \gamma_N, \overline{\Omega}_* \rangle \Omega_* = \gamma_N $ and $-\i \langle \gamma_S, \Omega_*\rangle \overline{\Omega}_* = \gamma_S$, the following identities hold
\begin{equation}
    \i \overline{Z}(H_N,\Omega_*) \Omega_*  \= H_N\,, \qquad -\i Z(H_S,\Omega_*) \overline{\Omega}_*\= H_S \,.
\end{equation}
Adding these two equations and using \eqref{rere} shows that our solution satisfies the stabilization equation \eqref{sasa} everywhere. To determine the metric, we first compute $\Sigma(H)$. After setting ${\sf h}_S ={\sf h}_N = 1/ |Z_*(\Gamma)|$ and recalling that $Z_*(\gamma_S)=Z_*(\Gamma)$ and $\overline{Z}_*(\gamma_N)=\overline{Z}_*(\Gamma)$, the emblackening factor becomes
\be
    \Sigma(H) \= \langle H, \Omega_* \rangle \langle H, \overline{\Omega}_*\rangle =   \langle H_S, \Omega_* \rangle \langle H_N, \overline{\Omega}_*\rangle 
    \= \underbrace{\left(  1 +  \frac{ |Z_*(\Gamma)|}{|\xvec-\xvec_S|} \right)}_{\widetilde{V}(\xvec)}\underbrace{ \left(1 +  \frac{|Z_*(\Gamma)|}{|\xvec-\xvec_N|} \right)}_{V(\xvec)}\,,\label{onon}
\ee
which manifestly satisfies $\Sigma(|\xvec|\to\infty) = 1$. As indicated in the second line, this emblackening factor becomes precisely the one that appears in the IWP solution \eqref{eq:IWPmetric}, with the black hole charge replaced by the magnitude of the central charge $Q\to |Z_*(\Gamma)|$. There remains to find the one-form $\omega$. Since $\langle \gamma_a, \gamma_a \rangle=0$ only the crossed terms of $\langle \dd H, H \rangle $ survive 
\begin{equation}\label{omom}
    \bast \dd \omega_E \=\i \langle \dd H_N , H_S \rangle + \i\langle \dd H_S , H_N\rangle \= \frac{\i\langle \gamma_N, \gamma_S \rangle}{\langle \gamma_N,\overline{\Omega}_* \rangle \langle \gamma_S, \Omega_*\rangle} \left( \widetilde{V}\dd V -V \dd \widetilde{V}\right)\,, 
\end{equation}
where we used the explicit value of ${\sf h}_a$. The ratio in the right hand side cancels thanks to the new attractor equations combined with $\langle \Omega_*, \overline{\Omega}_* \rangle = -\i$, namely
\begin{equation}
    \frac{\i\langle \gamma_N, \gamma_S \rangle}{\langle \gamma_N,\overline{\Omega}_* \rangle \langle \gamma_S, \Omega_*\rangle} \= \frac{\i \langle \i \overline{Z}_*(\gamma_N)\Omega_*, -\i Z_*(\gamma_S)\overline{\Omega}_*\rangle}{\overline{Z}_*(\gamma_N) Z_*(\gamma_S)} \=1 \,.
\end{equation}
It is now evident that equation \eqref{omom} reduces precisely to the IWP relation \eqref{eq:omega_equation}. One can verify a similar simplification in the electromagnetic field strength. As a final consistency check we can see that for the choices of ${\sf h}_N$ and ${\sf h}_S$ above we get $h=\Gamma/|Z_*(\Gamma)|$ which is consistent with the moduli at infinity being equal to the (old) attractor values for a black hole of charge $\Gamma$. We therefore conclude that for the boundary conditions specified by our choice of $h$ the black hole reduces to the IWP solution of Section \ref{sec:simplest-saddle-grav-index}. This can be generalized to a multicenter situation, but we leave it for future work \cite{wip}. 

\smallskip

We would like to make the following observation. Earlier, around equation \eqref{Zcenter}, we pointed out the possibility of implementing smoothness of our solution by imposing $\langle \gamma_a , \Omega_* (\gamma_a) \rangle = 0$ for $a=N, S$. If such a solution exists (which we rule out by other means in Section~\ref{sec:more-general-attractor}) it would not become of the IWP form appropriate to describe the topology of a black hole in any limit. To see this we can evaluate the emblackening factor
\begin{equation}
\label{eq:Sigma-H-Taub-NUT}
    \Sigma(H) = \langle h, \Omega_* \rangle \left( \langle h,\overline{\Omega}_* \rangle + \sum_a \frac{\langle \gamma_a , \overline{\Omega}_* \rangle}{|\xvec-\xvec_a|}\right).
\end{equation} 
This is similar to setting, in the IWP language, $V$ to be a constant and $\widetilde{V}$ to be a general harmonic function with two poles. Such a solution does not satisfy the smoothness condition appropriate to a black hole solution which imposes that the coefficient of the pole at $|\xvec|\to \infty$ for $V$ and $\widetilde{V}$ have to be the same (instead it has the topology of Taub-NUT). Therefore we conclude that only solutions with $\langle \gamma_N, \Omega_* \rangle = \langle \gamma_S , \overline{\Omega}_*\rangle = 0$ are connected to IWP and moreover, the fact the scalar takes the extremal attractor values implies real $n^I$ and $m_I$, through essentially the same argument in equations \eqref{Poleattrcomp}-\eqref{SPattrcomp}.

\smallskip

Finally, even though we leave a thorough study of the finite temperature multicenter configuration for future work \cite{wip} we can make some straightforward preliminary comments in relation to the IWP reduction. Given the single center attractor data, namely how the total charge $\Gamma$ is divided into $\gamma_N$ and $\gamma_S$, we can construct the harmonic vector $H=H_N + H_S$ with
\begin{equation}
H_N \= \gamma_N \left( {\sf h} + \sum_{a}  \frac{{\sf e}_a}{|\xvec-\xvec_a|}\right)\,, \qquad H_S \= \gamma_S \left(\widetilde{{\sf h}} + \sum_{a}  \frac{\widetilde{{\sf e}}_a}{|\xvec-\tilde{\xvec}_a|}\right) \,.   
\end{equation}
The positions $\xvec_a$ are a generalization of $\xvec_N$ while $\widetilde{\xvec}_a$ of $\xvec_S$. An obvious extension of the previous argument shows that this choice of $H$ together with the moduli being constant throughout space $(\Omega_*,\overline{\Omega}_*)$ gives a solution to the stabilization equations. To satisfy the metric boundary condition far away from the black hole we can set ${\sf h}_a = \widetilde{{\sf h}}_a = 1/|Z_*(\Gamma)|$. This leads to an emblackening factor 
\begin{equation}
    \Sigma(H) \= \left( 1 +\sum_a  {\sf e}_a \frac{|Z_*(\Gamma)|}{|\xvec - \xvec_a|}\right)\left( 1 +\sum_a \widetilde{{\sf e}}_a \frac{|Z_*(\Gamma)|}{|\xvec - \widetilde{\xvec}_a|}\right) \,.
\end{equation}
Just like the first case we considered, we see that ${\sf e}_a$ and $\widetilde{{\sf e}}_a$ can be interpreted as the fraction of the total charge $|Z_*(\Gamma)|$ distributed among each center. If we call $V(\xvec)$ the first term on the right hand side and $\widetilde{V}(\xvec)$ the second, the one-form $\omega$ satisfies the same equation as~\eqref{eq:omega_equation}. 
Thus we reproduce  
the multicenter generalization of the IWP metric from the most general solution of $\mathcal{N}=2$ supergravity when the moduli are set to their attractor value at infinity. At this point the smoothness analysis carried out in \cite{Hartle:1972ya,Yuille:1987vw,Whitt:1984wk} can be transferred to our problem.

\smallskip

We have seen in this section that the same boundary condition for the moduli at infinity that simplifies the extremal $\mathcal{N}=2$ black hole into extremal Reissner-Nordstrom also simplifies the rotating black hole at finite temperature into supersymmetric Kerr-Newman, and the same works in the multicenter case. This fact was implicitly used in \cite{H:2023qko}, which used the IWP solution to compute the index of black holes in supergravity theories involving vector multiplets. Such a simplification can also be useful when trying to extend the localization analysis of \cite{Iliesiu:2022kny}, done in the near horizon region, to the full asymptotically flat space geometry, since the Kerr-Newman solution is much simpler than the general case.

\subsection{The attractor solution as the unique complex saddle for the index}

\label{sec:more-general-attractor}

Above, we have assumed that the saddle for the index comes from an analytic continuation of a real Lorentzian solution. Without making this assumption, and instead looking at all possible complex Euclidean solutions, we will now argue that the attractor solution described above is, in fact, the unique saddle that satisfies the necessary boundary conditions for an index. Even without the assumption of analytic continuation, the stabilizer equations \eqref{genstabEuc} still need to be satisfied. After imposing that there are no higher order poles in $e^{-2U}$, \eqref{genstabEuc} implies that one of the following two conditions needs to be satisfied,\footnote{As before, one could also have analogous relations 
with the roles of $Z_*(\gamma_N)$ and~$\overline{Z}_*(\gamma_N)$ reversed which will yield solutions related by complex conjugation. } 
\be \label{Poleattr1}
\textit{Option I. }\,\,\,\,\text{North pole: }\,\,\,  \i\overline{Z}_*(\gamma_N) \,\Omega_*(\gamma_N) \= \gamma_N  \,,\quad 
\text{South pole: } \,\,\,-\i Z_*(\gamma_S) \, \overline{\Omega}_* (\gamma_S) \=  \gamma_S  \,, 
\ee
\be \label{Poleattr2}
\textit{Option II. }\,\,\,\,\text{North pole: } \,\,\,  \i\overline{Z}_*(\gamma_N) \,\Omega_*(\gamma_N) \= \gamma_N  \,,\quad 
\text{South pole: } \,\,\, \i\overline{Z}_*(\gamma_S) \,\Omega_*(\gamma_S) \= \gamma_S  \,,
\ee
where we should now study solutions in which $(n^I, m_I)$ can have components with both real and imaginary parts.  

First, let us analyze the consequences of Option I, which is the condition also satisfied by the solution described above whose analytic continuation to Lorentzian signature is smooth. In components, by taking the real and imaginary parts of \eqref{Poleattr1} and by rewriting the equations in terms of the scale-invariant variables $\overline{Z}(\gamma_N)\Omega(\gamma_N)= (Y^I_N, G_{IN})$ and $Z(\gamma_S) \overline{\Omega} (\gamma_S) = (\overline{Y}^I_S, \overline{G}_{IS})$
, we find
\be \label{stabnew-1}
\begin{split}
\i (Y^I_N- (Y^I_N)^*)   \= P^I - \Im \,n^I \,, & \qquad  
\i (G_{IN}- G_{IN}^*) \= Q_I - \Im \, m_I \,, \\
Y^I_N+ (Y^I_N)^*  \= \Re\,n^I \,, & \qquad  
G_{IN}+ G_{IN}^* \= \Re\, m_I \,,
\end{split}
\ee
for the North Pole,
while at the South Pole we have 
\be \label{stabnew-2}
\begin{split}
\i((\overline{Y}^I_S)^*- (\overline{Y}^I_S))   \= P^I + \Im \,n^I \,, & \qquad  
\i(\overline{G}_{IS}^*- \overline{G}_{IS}) \= Q_I + \Im \, m_I \,, \\
(\overline{Y}^I_S)^*+ \overline{Y}^I_S  \= \Re\,n^I \,, & \qquad  
\overline{G}_{IS}^*+ \overline G_{IS} \= \Re\, m_I \,.
\end{split}
\ee
From these equations, we would like to solve for the charges $(n^I, m_I)$ and for the values of the scalars at the north and south pole. There are $8n_V+8$ real equations with $8n_V+8$ real unknowns.\footnote{The unknowns are $\Re\, Y_N^I$, $\Im \,Y_N^I$, $\Re\, \overline Y_{S}^I$, $\Im\, \overline Y_{S}^I$, $\Re \,n^I$, $\Im\, n^I$, $\Re\, m_I$, $\Im \, m_I$, where we have used the definition of $G_{IN}$ and $\overline G_{IS}$ are functions of $Y_N^I$ and $\overline Y_{S}^I$, respectively.}  The first lines of \eqref{stabnew-1} and \eqref{stabnew-2} are equivalent to the old attractor equations \eqref{stab} for two different attractor black holes, one with real charges $\Re \,\gamma_N =\frac{1}2 (P^I-\Im\, n^I, \, Q_I - \Im \, m_I)$ and the other with real charges  $\Re \,\gamma_S = \frac{1}2(P^I+\Im\, n^I, \, Q_I + \Im \, m_I)$. These two equations fully determine $(Y_N^I, G_{IN})$ and $(\bar Y_S^I, \bar G_{IS})$ as a function of $\Re\, \gamma_N$ and $\Re \,\gamma_S$. 

By solving these equations, one finds the real part of the scalars at each pole is given by 
\be 
2(\Re \,Y^I_N,\,\Re\, G_{IN} )^\alpha = I^{\alpha \beta}\left(\frac{\partial \Sigma(\gamma)}{\partial \gamma^\beta}\right)_{\gamma={\rm Re}\,\gamma_N},  \,\, 2(\Re \,\bar Y^I_S, \Re\, \bar G_{IS})^\alpha = I^{\alpha \beta}\left(\frac{\partial \Sigma(\gamma)}{\partial \gamma^\beta}\right)_{\gamma={\rm Re}\,\gamma_S}, 
\ee
where $\alpha = 1,\,\dots,\,2n_V+2$. The function $\Sigma(\gamma) = Z_*(\gamma) \bar Z_*(\gamma)$ is obtained from the attractor solution in the first lines \eqref{stabnew-1} and \eqref{stabnew-2}. The non-trivial constraint that still needs to be satisfied comes from the second line of \eqref{stabnew-1} and \eqref{stabnew-2} by imposing that the real parts of all scalars agree, 
\be 
\Re\,Y_N^I = \Re \,\overline{Y}_S^I\,,\qquad \Re \,G_{IN} = \Re \,\overline{G}_{IS}
\ee 
which, in turn, implies
\be 
\label{eq:what-we-still-need-to-solve}
 \left(\frac{\partial \Sigma(\gamma)}{\partial \gamma^\alpha}\right)_{\gamma={\rm Re}\,\gamma_N} = \left(\frac{\partial \Sigma(\gamma)}{\partial \gamma^\alpha}\right)_{\gamma={\rm Re}\,\gamma_S}\,, 
\ee
We have thus reduced the $8n_V+8$ equations with $8n_V+8$ unknowns, \eqref{stabnew-1} and \eqref{stabnew-2}, to \eqref{eq:what-we-still-need-to-solve}, a system of $2n_V+2$ equations with $2n_V+2$ unknowns, $(\Im\, n^I,\, \,\Im\, m_I)$. This system can now be solved for specific prepotentials. 

For concreteness, let us focus on the 
theory with one vector multiplet and with the prepotential~\eqref{eq:diagonal-prepotential}, that we discussed in our Calabi-Yau example, for which~$\Sigma(\gamma)$ takes the form \eqref{eq:Sigma(H)-diagonal-prepotential}. 
We can always go to the duality frame in which two of the total charges are non-zero, as in the D4-D0 solution, $\Gamma= (0,\hat{p},0,\hat{u})$.
Numerically, we find that the only solution to~\eqref{eq:what-we-still-need-to-solve} is $\Im\,n^I = \Im\,m_I = 0$. The dipole charges are, therefore, uniquely fixed by the real parts of the scalars to be 
\be
m_I \= G_{I*} + \overline{G}_{I*}  \,, \qquad n^I \= Y^I_* + \overline{Y}^I_*\,.
\ee
A duality transformation can then be  used to infer  that the same is true for the solution with any~$\Gamma$. 
After imposing that time is periodic, we recover the attractor solution from Section~\ref{sec:new-attractor}.

To show the uniqueness of the attractor solution, what is left is to analyze the second option \eqref{Poleattr2}. 
In components, the equations at the north pole \eqref{stabnew-1} remain unchanged, but the equations at the south pole become 
\be 
\label{stabnew-3}
\begin{split}
\i(Y^I_S- (Y^I_S)^*)   \= P^I + \Im \,n^I \,, & \qquad  
\i(G_{IS}- G_{IS}^*) \= Q_I + \Im \, m_I \,, \\
Y^I_S+ (Y^I_S)^*  \= -\Re\,n^I \,, & \qquad  
G_{IS}+ G_{IS}^* \= - \Re\, m_I \,,
\end{split}
\ee
We again obtain a system of $8n_V+8$ equations with $8n_V+8$ unknowns. Pursuing the same strategy as above, \eqref{stabnew-3} can also be reduced to a system of $2n_V+2$ equations
\be 
\label{eq:what-we-still-need-to-solve-2}
 \left(\frac{\partial \Sigma(\gamma)}{\partial \gamma_I}\right)_{\gamma={\rm Re}\,\gamma_N} = -\left(\frac{\partial \Sigma(\gamma)}{\partial \gamma_I}\right)_{\gamma={\rm Re}\,\gamma_S}\,, 
\ee
which again has $2n_V+2$ unknowns, $(\Im\, n^I,\, \,\Im\, m_I)$. 
Focusing once again on the 
prepotential~\eqref{eq:diagonal-prepotential} that we discussed in our Calabi-Yau example 
with~$n_V=1$, we find no solutions for which the symplectic vector  $(\Im\, n^I,\, \,\Im\, m_I)$ only has real components. Thus, option II does not yield any additional smooth solutions for total charges that are real.\footnote{Option II could be additional solutions in the case where $Q^I$ and $P^I$ are complex. This should yield the    Taub-NUT solutions discussed in \eqref{eq:Sigma-H-Taub-NUT} which typically have complex charges.  }

Thus, at least for the theory \eqref{eq:diagonal-prepotential}, the new attractor solution is the unique complex solution that satisfies index boundary conditions. For any prepotential, \eqref{eq:what-we-still-need-to-solve} and \eqref{eq:what-we-still-need-to-solve-2} provide a simple system of equations that can be solved exactly and always has the new attractor saddle with $\Im\,n^I = \Im\,m_I = 0$ as a solution. It would be interesting to understand whether or not there are prepotentials or models with additional vector multiplets for which additional complex saddles can be found   (that still correspond to single-center black hole contributions).

\section{On-shell action and the free energy of the index}

\label{sec:on-shell-action-and-free-energy-of-the-index}

The above discussion of the new attractor mechanism completes the analysis of the regularity of the new finite temperature supersymmetric solutions. We now turn our attention to their contribution to the gravitational path integral that computes the supersymmetric index. For that, we will evaluate the on-shell action of the 
solution \eqref{eq:Bates-Denef_metric_Euclidean}. As we will see, the regularity conditions play a key role in deriving the final answer.

The bulk part of the Wick rotated action \eqref{eq:Lorentzian_action} is 
\begin{align}
-S_{\text{bulk}} &\= \frac{1}{16\pi} \int \dd^4 x \sqrt{g} R 
- \int G_{A\bar{B}} \dd t^A \wedge \ast \dd \bar{t}^B 
- \frac{\i}{16\pi} 
\int F^I \wedge G_I
\,,
\\ 
-\i \int F^I \wedge G_I &\= 
\int (
\Im \CN_{IJ}\, F^I \wedge \ast F^J 
+ \i\, \Re \CN_{IJ}\, F^I \wedge F^J 
) \,,
\end{align}
where the Hodge star $\ast$ is now taken with respect to Euclidean metric. 
The dual field $G_I$ is now
\be 
G_I \= \i\, \Im \CN_{IJ} \, \ast F^J - \Re \CN_{IJ} \,F^J \,. 
\ee
In addition, the total action is supplemented by the following two boundary terms 
\begin{align}
-S_{\text{GHY}}^{\text{boundary}} &\=
\frac{1}{8\pi} \int \dd^3 x \sqrt{h} K|_{\text{reg}}
\,, 
\\ 
-S_{\text{EM}}^{\text{boundary}} &\= 
\frac{\i}{16\pi}
\int \dd^3 x \sqrt{h}\, n_\mu \frac{1}{\sqrt{g}} \epsilon^{\mu \nu\rho \sigma} G_{I, \rho \sigma} A^I_\nu 
\label{eq:EM_bdry_term}
\,,
\end{align}
where the first term is the standard Gibbons-Hawking-York term, while the second term is required when working in the ensemble with fixed electric charge at infinity. The total action we want to evaluate on the finite temperature solutions is then 
\be 
-S_{\text{total}} \= -S_{\text{bulk}}  -S_{\text{GHY}}^{\text{boundary}}  -S_{\text{EM}}^{\text{boundary}} 
\,.
\ee
Let us now analyze separately contributions coming to the total on-shell action from different fields.

\paragraph{Gravity \& scalar contributions:} To evaluate the bulk part of the action containing the Ricci scalar and the scalar fields, it is simplest to use the trace of the Einstein equations following from the above action. Because the electromagnetic stress tensor in four dimension is traceless, the resulting equation will fix the Ricci scalar in terms of the scalar fields. One finds that its value is precisely such that 
\be 
\int \dd^4 x \sqrt{g} \left( \frac{R}{16\pi} - G_{A\bar{B}} \partial_\mu t^A 
\partial_\nu \bar{t}^B \right)_{\text{on-shell}} = \; 0.
\ee
Next we turn our attention to the Gibbons-Hawking boundary term. Here we encounter a divergent piece, which we regularize using a standard procedure of subtracting a vacuum contribution with the same boundary metric 
\be 
\sqrt{h} K|_{\text{reg}} \;\equiv\; \sqrt{h} (K - K_0) .
\ee
The resulting regulated Gibbons-Hawking-York boundary term gives then a simple contribution proportional to the ADM mass of the black hole
\be 
\frac{1}{8\pi} \int \dd^3 x \sqrt{h} K|_{\text{reg}} \= 
\frac{1}{8\pi} \int \dd\tau \dd\theta \dd\phi \, r_0^2 \sin\theta \frac{\partial_r \Sigma}{2 \Sigma} 
\= - \frac{\beta}{2} |Z(\Gamma;\Omega_\infty)|
\,.
\ee

\paragraph{Gauge fields:} To evaluate the contribution coming from the gauge fields, it will be convenient to write the field strength $\mathcal{F} = \dd \mathcal{A} = (F^I,G_I)$ in terms of electric potentials $\Phi=(\tPhi^I , \Phi_I)$ and magnetic potentials $\chi=(\tilde{\chi}^I , \chi_I)$ as
\be 
\mathcal{F} \= 
\i \dd\Phi \wedge \bigl( \dd t_E + \omE \bigr)
+ \Sigma \bast \dd \chi ,
\ee
where the potentials are explicitly given by
\be 
\Phi^\alpha \= - I^{\alpha \beta} \partial_{\Hv^{\beta}} \log \Sigma \,, 
\qquad 
\dd\chi^\alpha \= \frac{1}{\Sigma} 
(-\Phi^\alpha \langle \dd\Hv, \Hv \rangle +
\dd\Hv^\alpha
) \,,
\ee
and we have introduced a shorthand notation $\partial_{\Hv^{\beta}} \log \Sigma \equiv \partial_{\Hv^{\beta}} \log \Sigma(\Hv)$.
In terms of these fields, the field strength takes a particularly nice form in the tetrad basis ($\mathcal{F}_{\mu \nu} \dd x^\mu \wedge \dd x^\nu = \tilde{\mathcal{F}}_{a b} e^a \wedge e^b$)
\be 
\tilde{\mathcal{F}}^\alpha_{0i} \= 
-\i \partial_i \Phi^\alpha , 
\qquad 
\tilde{\mathcal{F}}^\alpha_{ij} \=
\epsilon_{ijk} \partial_k \chi^\alpha \,.
\ee
With this, we now want to evaluate the expression
\begin{align} 
\i \int F^{I} \wedge G_{I} 
\= 
\int \dd^4 x \sqrt{g} 
\left(
\partial_i \tPhi^{I} \partial_i \chi_{I} 
+ \partial_i \Phi_{I} \partial_i \tilde{\chi}^{I}
\right) 
\,.
\end{align}
Inserting the expression for the magnetic potential and integrating by parts, the integral separates into asymptotic and bulk pieces
\begin{align}
\i \int F^{I} \wedge G_{I}  &\= \int \dd^4 x \, \partial_i (\tPhi^I \partial_i H_I +  \Phi_I \partial_i \tH^I  -  \tPhi^I \Phi_I \langle \partial_i \Hv ,\Hv \rangle) 
\\ 
&\phantom{\=} + \int \dd^4 x (
 \tPhi^I \Phi_I \langle \partial_i^2 \Hv ,\Hv \rangle 
-\tPhi^I \partial_i^2 H_I -  \Phi_I \partial_i^2 \tilde{H}^I ) 
\,.
\label{eq:EM_bulk_two_pieces}
\end{align}
To evaluate the asymptotic parts, we introduce the notation $\phi \equiv \Phi (r \rightarrow \infty )$ and use that $\langle \Gamma ,h \rangle =0$. This results in
\begin{align}
 \int \dd^4 x \; \partial_i (\tPhi^I \partial_i H_I +  \Phi_I \partial_i \tilde{H}^I  -  \tPhi^I \Phi_I \langle \partial_i \Hv ,\Hv \rangle) &
\= - 4\pi \beta (\tphi^I Q_I + \phi_I P^I ) 
\, .
\end{align}
For the bulk pieces, because $\partial_i^2 H = -4\pi [\gamma_N \delta(\xvec - \xvec_N)+
\gamma_S \delta(\xvec - \xvec_S)]$, each term localizes to contributions coming purely from the poles. We therefore need to find the values of the electric potential at the north pole and the south pole. These values get fixed in terms of the charges and the temperature by both the regularity condition \eqref{eq:regularity_condition_part1} and the new attractor equations \eqref{eq:new_attractor_charges}. 
To see this, we use the definition of $\Sigma(\Hv)$ to write ($\Omega_\alpha \equiv I_{\alpha \beta} \Omega^\beta$)
\be 
\frac{\partial \Sigma}{\partial \Hv^\alpha} \= \overline{Z}(\Hv) \Omega_\alpha(H) + Z(\Hv) \overline{\Omega}_\alpha(\Hv) \,.
\label{eq:Sigma_derivative}
\ee
Zooming in onto north and south poles, we obtain for derivatives of the entropy function
\be 
\frac{\partial \Sigma}{\partial H^\alpha}|_{\rho_N} \= \frac{1}{\rho_N} 
\overline{Z}(\gamma_N) \Omega_\alpha(\gamma_N) 
\,, 
\qquad 
\frac{\partial \Sigma}{\partial H^\alpha}|_{\rho_S} \= \frac{1}{\rho_S} 
Z(\gamma_S) \overline{\Omega}_\alpha(\gamma_S) 
\,,  
\ee
while for the entropy function itself we get
\be 
\Sigma(\Hv)|_{\rho_N} 
\; \simeq \;
\frac{1}{\rho_N} \overline{Z}(\gamma_N) \langle \Hv(\xvec_N) , \Omega(\gamma_N ) \rangle 
\,,
\qquad
\Sigma(\Hv)|_{\rho_S} 
\; \simeq \;
\frac{1}{\rho_S} Z(\gamma_S) \langle \Hv(\xvec_S) , \overline{\Omega}(\gamma_S ) \rangle
\,.
\ee
Combining the above expressions and using \eqref{eq:regularity_condition_part1} together with \eqref{eq:new_attractor_charges}, the values of the electric potentials at the poles get fixed to 
\be 
\Phi (\xvec_N) \= \frac{4\pi \i}{\beta} \gamma_N , 
\qquad 
\Phi (\xvec_S) \= -\frac{4\pi \i}{\beta} \gamma_S .
\ee
With this, 
the bulk parts of \eqref{eq:EM_bulk_two_pieces} evaluate to 
\be 
 \int \dd^4 x (
 \Phi^I \Phi_I \langle \partial_i^2 H ,H \rangle 
-\Phi^I \partial_i^2 H_I -  \Phi_I \partial_i^2 H^I ) 
\=
8\pi^2 \i (\i m_I P^I +  \i n^I Q_I)
\,.
\ee
Putting both pieces together, the bulk electromagnetic term \eqref{eq:EM_bulk_two_pieces} gives the contribution
\be 
- \frac{\i}{16\pi}  \int F^I \wedge G_I \= 
\frac{\beta}{4} 
( \tphi^I Q_I   + \phi_I P^I ) 
- \frac{\pi \i}{2} (\i m_I P^I +\i n^I Q_I) \, .
\ee
Lastly, we need to include to boundary electromagnetic term \eqref{eq:EM_bdry_term}. This term evaluates to 
\be 
-S_{\text{EM}}^{\text{boundary}} \=
- \frac{\beta}{2} Q_I \tphi^I 
+  \pi \i (\i n^I Q_I)
\,.
\ee
Its effect will be to change the signs of terms proportional to electric monopole charges $Q_I$. The total electromagnetic contribution to the on-shell action is therefore 
\begin{align}
- \frac{i}{16\pi}  \int F^I \wedge G_I  -S_{\text{EM}}^{\text{boundary}} 
&\= 
\frac{\beta}{4} ( 
-\tphi^I Q_I   + \phi_I P^I ) 
- \frac{\pi \i}{2} (\i m_I P^I - \i n^I Q_I) 
\\ 
&\= 
-\frac{\beta}{4} \Gamma^\alpha \partial_{\Hv^\alpha} \Sigma(\Hv)|_\infty + \pi \i \langle \gamma_N ,\gamma_S \rangle. 
\label{eq:EM_total_part1}
\end{align}
Let us take a closer look at the terms appearing in the above expression.
The linear combination appearing in front of the inverse temperature $\beta$ is nothing but the ADM mass. To see this, we again make use of the relation \eqref{eq:Sigma_derivative} to find
\begin{align}
\Gamma^\alpha \partial_{\Hv^\alpha} \Sigma(\Hv)|_\infty \= \langle h , \overline{\Omega}_\infty \rangle 
\langle \Gamma, \Omega_\infty \rangle   
+ 
\langle h , \Omega_\infty \rangle
\langle \Gamma, \overline{\Omega}_\infty \rangle 
\= 2 |Z(\Gamma;\Omega_\infty)| 
\,.
\end{align}
For the second term appearing in \eqref{eq:EM_total_part1}, we insert the expression for the charges fixed by the new attractor equations \eqref{eq:new_attractor_charges}
\be 
\pi \i \langle \gamma_N , \gamma_S \rangle
\= 
\pi  Z_*(\Gamma)  \overline{Z}_* (\Gamma) 
\=\pi \Sigma(\Gamma) 
\,, 
\ee
which is exactly the entropy of the extremal black hole $\SBH$ with total charge $\Gamma$, as given in~\eqref{SBH2}.

With this, the total contribution of the gauge fields to the on-shell action can be written as
\be 
- \frac{\i}{16\pi}  \int F^I \wedge G_I  -S_{\text{EM}}^{\text{boundary}} \=
-\frac{\beta}{2} |Z(\Gamma;\Omega_\infty)|
+ \pi \Sigma(\Gamma)
\, . 
\ee
\paragraph{Total on-shell action:} 
Putting the above contributions together,
we obtain the total Euclidean on-shell action 
\be 
-S_{\text{total}} 
\= -\beta \underbrace{|Z(\Gamma;\Omega_\infty)|}_{=E_{BPS}} \, + \, 
\underbrace{\pi \i
\langle \gamma_N , {\gamma}_S \rangle}_{\substack{\=\log(d_b-d_f) \\ \= \SBH} } 
\,.
\ee
This is precisely the expected answer for a supersymmetric index. The first term captures the contribution of the BPS energy to the logarithm of the index and is given by the central charge $|Z(\Gamma;\Omega_\infty)|$. The second term is expected to capture the logarithm of the difference between the bosonic BPS degeneracy and the fermionic BPS degeneracy. While this term is different than the area of the black hole solutions that we study, which instead has non-trivial dependence on $\beta$ and on the moduli values, it is exactly the same as the extremal entropy of the classical attractor solution with the same set of electric and magnetic charges. 

\paragraph{The free energy of the index: } Similarly, as in the D0-D4 example from the previous section, we can see the cancellation of the temperature dependence already at the level of free energy. Writing down the first regularity condition 
\be 
\frac{ \i \langle \gamma_N , \gamma_S \rangle}{|\xvec_N - \xvec_S|} + \i \langle \gamma_N ,h \rangle \= 
\frac{\beta}{4\pi} \,,
\ee
we can again rewrite it as
\be 
\beta \Omega J + \frac{A_{\text{H}}}{4} = \pi \i \langle \gamma_N , \gamma_S \rangle \,. 
\ee
By interpreting the area term as the finite temperature entropy of the solution and inserting the new attractor charges \eqref{eq:new_attractor_charges} we find
\be 
\beta \Omega J + \frac{A_{\text{H}}}{4}\= \SBHext \,,  
\ee
which shows that the quantum statistical relation for the index is satisfied for our finite temperature solutions in the case with general charges and arbitrary number of vector multiplets. This simple computation also highlights that the final result is essentially fixed by regularity conditions imposed on the Euclidean geometry.

\section{Discussion}
\label{sec:discussion}

In this paper, we have found the saddles that contribute to the gravitational index, 
which capture the contributions of the BPS states associated with a single black hole. In contrast to the classical extremal black hole solutions of supergravity, the Gibbons-Hawking prescription for the gravitational index requires our solutions to have a finite temperature and to be rotating, nevertheless preserving supersymmetry. The on-shell action of these saddles is always a sum of two terms. The first term is temperature-independent and proportional to the extremal black hole area, implying it is also independent of asymptotic moduli. 
The second term is proportional 
to~$\beta$, therefore carrying the information about the BPS energy, and can depend on the asymptotic moduli. This is exactly the behavior expected for the logarithm of a supersymmetric index. 
The difference between bosonic and fermionic BPS degeneracies that we predict fully agrees with the extremal entropy of the classical attractor solution with the same asymptotic charges. 
This, in turn, is different from the area of the black hole solutions that we study, which instead has a complicated dependence on temperature and the moduli values. 
The simplicity of the on-shell action can be viewed as a consequence of the attractor equations applied to our smooth Euclidean solution: not only the on-shell action but also the value of the scalars and of the electric and magnetic fields in each vector multiplet are fixed at the north and south poles of the black hole to values that are solely dependent on the black hole's electric and magnetic charges.  

Our solutions can also be generalized to obtain other saddle-point contributions of the gravitational index. For instance, by introducing additional north and south pole pairs in our harmonic functions, we can find saddles that capture the contribution of multi-center bound states to the gravitational index.  Even though these solutions have a finite temperature, the multiple black holes can still exist in equilibrium due to their angular velocities. The various properties of these solutions, including smoothness, attractor flow, and the evaluation of their on-shell action, shall be analyzed 
in an upcoming paper \cite{wip}.

The saddles of the gravitational index, including the solutions described above, are purely Euclidean solutions and have no Lorentzian continuation with a smooth horizon.\footnote{The Euclidean saddles that appear in the evaluation of the index at finite temperature do not need to match the geometric description of the Lorentzian states in a canonical treatment. When we compute the index at finite temperature, we have rotating Euclidean saddles, even though we are counting states described by extremal supersymmetric black holes.} 
One issue is that the rules about which saddles to include in the Euclidean gravitational path integral are still not well-established.  
In particular, some of the saddle points analyzed above include analytic continuations of the fields in the vector and graviton multiplets, 
and so, in order for such saddles to contribute to the path integral, one has to deform the integration contour for some of these fields to include such saddles. Whether a contour deformation is possible such that 
a contour of steepest descent passes through this saddle is difficult to tell a priori. 
One can, however, take the opposite perspective. Assuming a UV completion of the gravitational path integral exists, and assuming such an object is consistent with the black hole holographic dual, we can infer which saddles should or should not contribute.\footnote{This does not rule out other possible UV completions of the gravitational path integral. Nevertheless, all examples known from String Theory are holographic and therefore one might expect the UV completion is unique and holographic.} For example, since the index in quantum mechanics should not continuously change with the temperature or moduli value, the black hole solutions we found are forced to contribute to the Euclidean path integral. Since some of the discussed saddles are complex, this could perhaps be used in order to improve the rules about which saddles contribute to the gravitational path integral (assuming holographic behavior). This perspective has already produced preliminary results in the context of supersymmetric black holes AdS${}_5$ and $\mathcal{N}=4$ Yang-Mills theory \cite{Aharony:2021zkr}.

 Our saddle point solutions can also be used to compute quantum corrections to the index. This program was already initiated in \cite{H:2023qko, Anupam:2023yns}, where quantum corrections around the Einstein-Maxwell saddlepoint presented in Section \ref{sec:simplest-saddle-grav-index} were computed. As we have explained in Section \ref{sec:some-examples}, such solutions correspond to the supergravity saddles in which the scalars are fixed to their attractor value on the entire spacetime but are not the most general solutions. It would therefore be instructive to compute quantum corrections around the saddles that have arbitrary values for their asymptotic moduli. This will allow for a better understanding of why the entire partition function, rather than only the on-shell action of the saddle, is independent of the temperature and of the asymptotic values of the moduli.\footnote{At low and high temperatures, the quantum corrections around such saddles are quite different. At low temperatures, when the black hole is near-extremal, there are large quantum corrections coming from the super-Schwarzian modes that capture fluctuations of the metric and gravitino in the near-horizon region \cite{Heydeman:2020hhw, Iliesiu:2021are, Boruch:2022tno, Iliesiu:2022onk}. At high temperatures, when the saddle is more akin to a Schwarschild solution, no such soft modes exist. Nevertheless, for the gravitational index, the quantum corrections in the two limits have to agree.}

 Our results also serve as the starting point for localizing the supergravity path integral with asymptotically flat boundary conditions. While there have been several developments in localizing the supergravity path integral with AdS$_2 \times S^2$ boundary conditions to reproduce the degeneracy of $1/8$-BPS black holes in string theory compactified on $\mathbb R^4 \times T^6$ \cite{Dabholkar:2010rm, Dabholkar:2010uh, Dabholkar:2011ec, Gupta:2012cy, Murthy:2013xpa, Dabholkar:2014ema, Gupta:2015gga, Murthy:2015yfa, Jeon:2018kec, Iliesiu:2022kny}, it would be useful to understand whether the same results can be reproduced in asymptotically flat space with the boundary conditions of a gravitational index. This would in particular address some of the concerns raised in \cite{Sen:2023dps} regarding the use of supersymmetric localization for these black holes. In that context, each one of our saddles will be a single point on the localization locus. The rest of the locus is obtained by solely imposing that the spacetime has a well-defined Killing spinor (which our solutions have) with respect to which the localization procedure is performed, but without imposing that all the supergravity equations of motion are obeyed. We hope to explore such a procedure in detail in the near future. 

In order to carry out the program outlined in the previous paragraph, there is one more generalization of the classical black hole solutions which needs to be elucidated. In this paper we focused on black hole solutions of $\mathcal{N}=2$ supergravity at the two-derivative level. Since the examples arising from String Theory come with specific higher derivative corrections (except for the compactification to $\mathcal{N}=8$ supergravity) we need to extend our analysis to prepotentials that lead to such corrections, namely $F(X) \to F(X,W^2)$ where $W^2$ is related to the square of the anti-self dual graviphoton field strength. In particular it would be important to derive the entropy formula for such black holes, 
see for example~\cite{Ooguri:2004zv} and references therein, purely from the gravitational path integral on the full asymptotically flat spacetime.   

Finally, it may be interesting to think about other possible applications of the solutions that we discuss in this paper. 
The construction of the Euclidean attractor solution as a Taub-NUT bubble is reminiscent of solutions of higher-dimensional supergravity where a similar bubble mechanism is used in Lorentzian solutions~\cite{Maldacena:2000dr,Lunin:2002iz,Warner:2019jll}.  
It would be interesting to understand if there are more physical relations between these constructions. 
In another direction, as mentioned in Section~\ref{sec:sugrarev}, one can interpret the classic attractor mechanism as a flow in the Calabi-Yau moduli 
space. The values of the moduli corresponding to the attractor points carry special number-theoretic properties~\cite{Moore:1998pn}. 
It would be interesting to understand if the new attractor mechanism discussed in this paper also has an interesting number-theoretic  interpretation.

\section*{Acknowledgements}

It is a pleasure to thank Pawel Caputa, Roberto Emparan, Chris Hull, David Katona, Maciej Kolanowski, James Lucietti,  Juan Maldacena, Boris Pioline, Joan Simon, and Marija Tomasevic for interesting and useful discussions. JB is supported by the NCN Sonata Bis 9 2019/34/E/ST2/00123 grant. LVI was supported by the Simons Collaboration on Ultra-Quantum Matter, a Simons Foundation Grant with No.~651440.
S.M.~acknowledges the support of the J.~Robert Oppenheimer 
Visiting Professorship at the Institute for Advanced Study, Princeton.
The work of GJT was supported by the Institute for Advanced Study and the NSF under Grant No.~PHY-2207584, 
and by the Sivian Fund, and currently by the University of Washington and the DOE award DE-SC0024363.
JB wants to thank King's College London for hospitality, where a part of this work has been done.

\bigskip

\appendix

\bibliography{EuclideanSusyBH}
\bibliographystyle{JHEP}

\end{document}